\documentclass[sigconf]{acmart}

\usepackage{subfigure}
\usepackage{graphicx}
\usepackage{cleveref}
\usepackage{makecell}

\usepackage{xcolor}

\newcommand{\gemmaTwoTwoB}{Gemma-2-2B}
\newcommand{\llamaThreeTwoOneB}{Llama-3.2-1B}
\newcommand{\llamaThreeTwoThreeB}{Llama-3.2-3B}
\newcommand{\mistralVZeroThreeSevenB}{Mistral-v0.3-7B}
\newcommand{\qwenTwoSevenB}{Qwen-2-1.5B}
\newcommand{\gemmaTwoNineB}{Gemma-2-9B}
\newcommand{\llamaThreeOneEightB}{Llama-3.1-8B}

\AtBeginDocument{%
  }

\setcopyright{acmlicensed}
\copyrightyear{2026}
\acmYear{2026}
\setcopyright{cc}
\setcctype{by}
\acmConference[WWW '26]{Proceedings of the ACM Web Conference 2026}{April 13--17, 2026}{Dubai, United Arab Emirates}
\acmBooktitle{Proceedings of the ACM Web Conference 2026 (WWW '26), April 13--17, 2026, Dubai, United Arab Emirates}
\acmPrice{}
\acmDOI{10.1145/3774904.3792508}
\acmISBN{979-8-4007-2307-0/2026/04}

\begin{document}

\title{Digital Skin, Digital Bias: Uncovering Tone-Based Biases in LLMs and Emoji Embeddings}

\author{Mingchen Li}
\affiliation{
  \institution{University of North Texas}
  \city{Denton}
  \state{TX}
  \country{United States}}
\email{MingchenLi@my.unt.edu}
\orcid{0009-0003-0704-5906}

\author{Wajdi Aljedaani}
\affiliation{
  \institution{International Center for Artificial Intelligence Research and Ethics}
  \city{Riyadh}
  \country{Saudi Arabia}}
\email{waljedaani@icaire.org}
\orcid{0000-0002-6700-719X}

\author{Yingjie Liu}
\affiliation{
  \institution{Southern Methodist University}
  \city{Dallas}
  \state{TX}
  \country{United States}}
\email{yingjiel@smu.edu}
\orcid{0009-0007-9711-8185}

\author{Navyasri Meka}
\affiliation{
  \institution{University of North Texas}
  \city{Denton}
  \state{TX}
  \country{United States}}
\email{NavyasriMeka@my.unt.edu}
\orcid{0009-0009-5073-4014}

\author{Xuan Lu}
\affiliation{
  \institution{The University of Arizona}
  \city{Tucson}
  \state{AZ}
  \country{United States}}
\email{luxuan@arizona.edu}
\orcid{0000-0003-4886-0918}

\author{Xinyue Ye}
\affiliation{
  \institution{The University of Alabama}
  \city{Tuscaloosa}
  \state{AL}
  \country{United States}}
\email{xye10@ua.edu}
\orcid{0000-0001-8838-9476}

\author{Junhua Ding}
\affiliation{
  \institution{University of North Texas}
  \city{Denton}
  \state{TX}
  \country{United States}}
\email{Junhua.Ding@unt.edu}
\orcid{0000-0002-2129-1586}

\author{Yunhe Feng}
\affiliation{
  \institution{University of North Texas}
  \city{Denton}
  \state{TX}
  \country{United States}}
\email{Yunhe.Feng@unt.edu}
\orcid{0000-0001-6577-227X}

\renewcommand{\shortauthors}{Mingchen Li et al.}

\begin{abstract}
Skin-toned emojis are crucial for fostering personal identity and social inclusion in online communication. As AI models, particularly Large Language Models (LLMs), increasingly mediate interactions on web platforms, the risk that these systems perpetuate societal biases through their representation of such symbols is a significant concern. This paper presents the first large-scale comparative study of bias in skin-toned emoji representations across two distinct model classes. We systematically evaluate dedicated emoji embedding models (emoji2vec, emoji-sw2v) against four modern LLMs (Llama, Gemma, Qwen, and Mistral). Our analysis first reveals a critical performance gap: while LLMs demonstrate robust support for skin tone modifiers, widely-used specialized emoji models exhibit severe deficiencies. More importantly, a multi-faceted investigation into semantic consistency, representational similarity, sentiment polarity, and core biases uncovers systemic disparities. We find evidence of skewed sentiment and inconsistent meanings associated with emojis across different skin tones, highlighting latent biases within these foundational models. Our findings underscore the urgent need for developers and platforms to audit and mitigate these representational harms, ensuring that AI's role on the web promotes genuine equity rather than reinforcing societal biases.
\end{abstract}



\maketitle

\section{Introduction and Background}
\label{sec:Introduction}

Emojis have been adopted extensively in various forms of web digital communication, such as social media, text messaging, emails, marketing and advertising, and even legal documents~\cite{feng2020new}. To satisfy the need for human diversity, skin-toned emojis that can represent important personal identity information (i.e., skin colors) have been proposed and approved by the Unicode Consortium\footnote{https://unicode.org/consortium/consort.html}.
Specifically, five skin tone modifiers \includegraphics[width=0.02\linewidth]{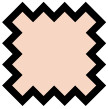},\includegraphics[width=0.02\linewidth]{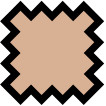},\includegraphics[width=0.02\linewidth]{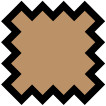},\includegraphics[width=0.02\linewidth]{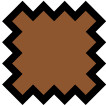},\includegraphics[width=0.02\linewidth]{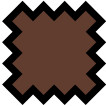}, designed based on the Fitzpatrick Scale~\cite{fitzpatrick1988validity}, were introduced to emoji characters in 2015 as part of Unicode 8.0\footnote{https://www.unicode.org/versions/Unicode8.0.0/} to enhance representation and inclusivity.
These modifiers are appended to emojis supporting skin tone adjustments, resulting in a change in the emoji's skin tone, e.g., \includegraphics[width=0.02\linewidth]{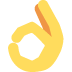} + \{\includegraphics[width=0.02\linewidth]{figures/emoji_image/light.jpg},\includegraphics[width=0.02\linewidth]{figures/emoji_image/med_light.jpg},\includegraphics[width=0.02\linewidth]{figures/emoji_image/med.jpg},\includegraphics[width=0.02\linewidth]{figures/emoji_image/med_dark.jpg},\includegraphics[width=0.02\linewidth]{figures/emoji_image/dark.jpg}\} $\rightarrow$ \{\includegraphics[width=0.02\linewidth]{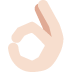},\includegraphics[width=0.02\linewidth]{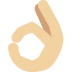},\includegraphics[width=0.02\linewidth]{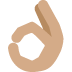},\includegraphics[width=0.02\linewidth]{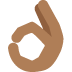},\includegraphics[width=0.02\linewidth]{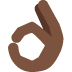}\}. More examples of emoji skin tone modifier sequences are illustrated in Table~\ref{tab:skin_toned_emoji_demo}.

\begin{table}[ht] 
\setlength{\tabcolsep}{6pt}
\centering
    \caption{Examples of skin-toned emojis}\label{tab:skin_toned_emoji_demo}
    \begin{tabular}{ccccccc} 
     \toprule
\textbf{Emoji} & \textbf{Unicode} &  \includegraphics[width=0.05\linewidth]{figures/emoji_image/light.jpg} & \includegraphics[width=0.05\linewidth]{figures/emoji_image/med_light.jpg} & \includegraphics[width=0.05\linewidth]{figures/emoji_image/med.jpg} & \includegraphics[width=0.05\linewidth]{figures/emoji_image/med_dark.jpg} & \includegraphics[width=0.05\linewidth]{figures/emoji_image/dark.jpg} \\
        \midrule
        \includegraphics[width=0.05\linewidth]{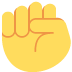} & U+270A & \includegraphics[width=0.05\linewidth]{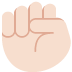} & \includegraphics[width=0.05\linewidth]{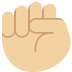} & \includegraphics[width=0.05\linewidth]{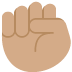} & \includegraphics[width=0.05\linewidth]{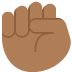} & \includegraphics[width=0.05\linewidth]{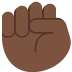} \\
        \includegraphics[width=0.05\linewidth]{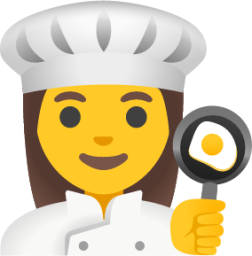} & U+1F44F & \includegraphics[width=0.05\linewidth]{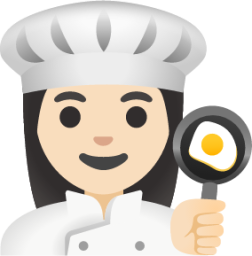} &\includegraphics[width=0.05\linewidth]{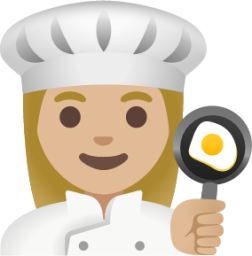} & \includegraphics[width=0.05\linewidth]{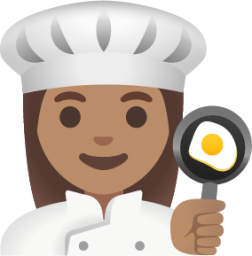} & \includegraphics[width=0.05\linewidth]{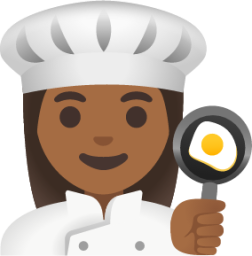} &  \includegraphics[width=0.05\linewidth]{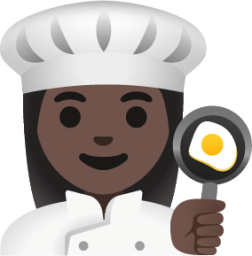} \\
        \includegraphics[width=0.05\linewidth]{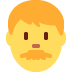} & U+1F468 & \includegraphics[width=0.05\linewidth]{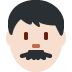} & \includegraphics[width=0.05\linewidth]{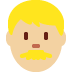} & \includegraphics[width=0.05\linewidth]{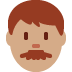} & \includegraphics[width=0.05\linewidth]{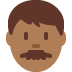} & \includegraphics[width=0.05\linewidth]{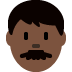} \\
        \includegraphics[width=0.05\linewidth]{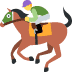} & U+1F3C7 & \includegraphics[width=0.05\linewidth]{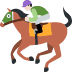} & \includegraphics[width=0.05\linewidth]{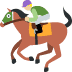} & \includegraphics[width=0.05\linewidth]{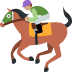} & \includegraphics[width=0.05\linewidth]{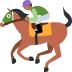} & \includegraphics[width=0.05\linewidth]{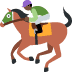} \\
     \bottomrule
    \end{tabular} 
\end{table}

According to the latest data from Emoji Counts (Version 17.0)\footnote{https://unicode.org/emoji/charts/emoji-counts.html}, approximately 51.36\% of the total 3,953 available emojis feature a skin-tone modifier. This functionality allows users to personalize emojis to better represent their own identities. Such emojis can be utilized independently to express attitudes or convey emotions, particularly in the context of web social media reactions to decisions or events. Furthermore, these skin-tone-specific emojis can be seamlessly integrated into textual content, including Twitter/X posts, to articulate individual viewpoints while simultaneously indicating their skin-tone identities. The integration of skin-tone emojis with text introduces multi-modality, enriching the conveyed information and adding diversity to the communication.

As language technologies increasingly mediate online interaction, the semantics assigned to tone‑modified emojis by pre‑trained models matter. Do skin‑tone variants receive \emph{consistent and equitable} treatment in representation and downstream associations? The question is non‑trivial for two reasons. First, widely used \emph{static} emoji/word embedding models were built before tone modifiers were prevalent, and thus may not support, or may under‑represent, toned variants. Second, while modern LLMs can compose dynamic embeddings via subword tokenizers, nothing guarantees that (i) their tokenization treats tones uniformly or (ii) their learned representations avoid sentiment and association skews. In short, support does not imply fairness.

To investigate this, we present the first large-scale comparative study of bias in skin-toned emoji representations across both dedicated emoji embedding models and modern LLMs. Our analysis unfolds in three stages. First, we establish a foundational performance baseline by benchmarking emoji support across nine prevalent static embedding models (e.g., emoji2vec~\cite{eisner2016emoji2vec}, emoji-sw2v~\cite{barbieri2018gender}) and four state-of-the-art LLMs. This reveals a critical gap: while LLMs offer comprehensive support for skin-toned emojis through sophisticated tokenizers, legacy static models show severe deficiencies, with most supporting five or fewer variants.

Second, we probe the semantic impact of skin tone modifiers within the models that support them. We employ a two-pronged approach to understand how skin tone variations influence meaning. To capture real-world usage, we analyze the similarity and clustering patterns of high-frequency emojis from emojitracker\footnote{\url{https://emojitracker.com/}}. To conduct a controlled experiment on affective polarity, we project emoji embeddings onto a ``Good'' vs. ``Bad'' (e.g., \includegraphics[width=0.02\linewidth]{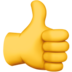} vs. \includegraphics[width=0.02\linewidth]{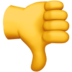}) semantic axis derived from emojidb\footnote{\url{https://emojidb.org/}}, quantifying any systematic drift toward positive or negative sentiment.

Third, forming the core of our contribution, we conduct a systematic quantification of the biases embedded within these representations. Drawing on established methodologies from text-based bias detection~\cite{caliskan2017semantics,hovy2021five}, we adapt and apply a suite of metrics to offer a multi-faceted analysis. We use Relative Norm Distance (RND)~\cite{garg2018word} to measure biases in a large group of hand emojis, the Word Embedding Association Test (WEAT)~\cite{caliskan2017semantics} and Relative Negative Sentiment Bias (RNSB)~\cite{sweeney2019transparent} to uncover associative biases in professional role emojis, and a standard WEAT benchmark to test associations between skin tones and societal concepts like gender, age, and race.

In this paper, we present a systematic analysis of skin tone representation in both static and dynamic emoji embedding models, leading to the following key contributions: 

\begin{itemize}
    \item We provide the first cross‑family benchmark of skin‑tone emoji support spanning nine static models and four LLMs, revealing substantial representational gaps in legacy systems.
    \item We introduce and quantify a tokenization‑level asymmetry for skin tones in LLMs—an overlooked axis of bias where certain tones (e.g., dark) incur higher token costs.
    \item Using real‑world high‑frequency emojis and a controlled affective axis, we show that tone modifiers can induce consistent semantic shifts even in modern LLMs.
    \item We adapt RND, WEAT, and RNSB to the emoji setting and deliver a multi‑faceted, statistically grounded measurement of associative and sentiment biases across tones, offering concrete evidence of representational harms.
\end{itemize}

Ultimately, we hope that these findings contribute to a nuanced understanding of how emojis, which signal specific demographic identities, should be processed and interpreted in web society.

\section{Related Work}
\label{sec:RelatedWork}

\subsection{Emoji and Social Media}
Prior studies have illustrated the multifaceted roles of emojis, encompassing their capacity to provide an emotional context within the text \cite{kaye2016turn} and convey or reinforce specific opinions \cite{hu2017spice}. It is essential to recognize that despite the potential of many emojis to substitute tangible entities they represent, their usage is not always straightforward. Interpretations of emojis tend to be personalized and can exhibit variations among individuals and small groups, often influenced by visual characteristics \cite{wiseman2018repurposing}. Additionally, studies have probed the paralinguistic and pro-social dimensions of emojis, with a particular emphasis on their prevalence in online expressions of solidarity during disaster-related events \cite{santhanam2019stand}.

Moreover, several studies have delved into how individuals perceive and comprehend emojis. A significant portion of these studies has concentrated on analyzing sentiment and affectivity, consistently finding that the majority of emojis tend to convey positive emotions. This consensus regarding the emotional impact of specific emojis holds true across different languages, as evidenced by Novak et al. \cite{kralj2015sentiment}. However, the potential for confusion arises from disparities in how emojis are displayed across diverse platforms, leading to varying interpretations of the same underlying emoji. This ambiguity can manifest whether emojis are used in isolation \cite{tigwell2016oh} or in conjunction with textual context \cite{miller2017understanding}.

\subsection{Technology Design and Bias in Emoji}
There is a substantial body of research highlighting the presence of inherent biases within technology, challenging the prevailing assumption of its neutrality. These biases tend to systematically favor specific groups, primarily white men, while simultaneously disadvantaging others, including women \cite{perez2019invisible}  and individuals from non-white backgrounds \cite{barbosa2019rehumanized, noble2018algorithms, raji2019actionable}. A notable manifestation of this bias can be observed in the representation of identity, where the default or dominant choice often aligns with whiteness \cite{brock2011keeping, cave2020whiteness, sparrow2020robotics}. It is worth noting that the development of technologies explicitly designed for non-white populations, as exemplified by Blackbird, a web browser tailored for Black internet users, often faces resistance. This resistance can be traced back to the prevailing belief that existing solutions inherently lack bias or partiality \cite{brock2011beyond}.

Similarly, the introduction of emoji skin-tone modifiers triggered a diverse array of responses. The original yellow emoji, made accessible in 2015, was viewed by some as neutral and inclusive of all individuals \cite{Arielle2023}. Consequently, doubts emerged regarding the practicality of introducing emoji variations featuring diverse skin tones \cite{zimmerman2015racially}. The inclusion of skin tones within the Unicode Standard has been a subject of contention, as evidenced by internal discussions within the Unicode Consortium, as elucidated by Miltner \cite{miltner2021one}. The sequencing of skin tones for encoding, whether in ascending or descending order of lightness, became a point of disagreement. Certain individuals expressed reservations about what they perceived as an excessive emphasis on correctness associated with a sequential arrangement from dark to light. Miltner's analysis suggests that ``the initial emoji selection process was influenced by a form of colorblind racism deeply embedded within institutional systems''. This type of racism often overlooks the significance of racial identity within ostensibly unbiased technological systems.

\section{Methodology}
\label{sec:Methodology}

\subsection{Skin-toned Emoji Identification}
\label{sec:emoji_check}
To establish a definitive set of skin-toned emojis for our analysis, we adhere to the specifications outlined in the Unicode Technical Standard\footnote{https://www.unicode.org/reports/tr51/}. This approach ensures that our emoji list is consistent with universal web and platform standards. Following this standard, we identify all emojis that incorporate one of the five Fitzpatrick scale skin tone modifiers \includegraphics[width=0.02\linewidth]{figures/emoji_image/light.jpg},\includegraphics[width=0.02\linewidth]{figures/emoji_image/med_light.jpg},\includegraphics[width=0.02\linewidth]{figures/emoji_image/med.jpg},\includegraphics[width=0.02\linewidth]{figures/emoji_image/med_dark.jpg},\includegraphics[width=0.02\linewidth]{figures/emoji_image/dark.jpg}. In our experiments, these are categorized as skin-toned emojis, while the base emojis without a modifier are explicitly labeled as `Default' and are not treated as a specific skin tone.

\subsection{Pre-trained Embedding Models}
\label{sec:pre-trained-models}

\subsubsection{Static Emoji Embedding Models}
\label{sec:static-models}
Static embedding models generate a single, fixed vector representation for each emoji, regardless of its surrounding context. These models typically operate with a predefined vocabulary, meaning they can only represent emojis included in their training data. All embedding vectors in this section exist in a $d$-dimensional space (i.e., $v \in \mathbb{R}^d$). In our analysis, we include two prominent examples:

\textbf{emoji2vec~\cite{eisner2016emoji2vec}} creates emoji embeddings by learning from their official Unicode descriptions. It aligns emojis with word vectors from a pre-trained word2vec~\cite{mikolov2013efficient} model by summing the vectors of the words in each emoji's description. This process maps emojis into the same 300-dimensional space as the word vectors, allowing for direct semantic comparison between words and emojis.

\textbf{emoji-sw2v~\cite{barbieri2018gender}} is trained on over 22 million tweets to produce a unified vector space for words, emojis, and their modifiers. Unlike models that treat modified emojis as single tokens, emoji-sw2v integrates modifiers directly into the neural network's architecture, aiming for a more nuanced semantic representation. However, like emoji2vec, its embeddings remain static once trained.

In addition to these two prominent models, our broader study includes general-purpose embedding models such as FastText~\cite{bojanowski2016enriching}, GloVe~\cite{pennington2014glove}, and BERT~\cite{devlin2018bert}, as well as other emoji dedicated models like emoji-word2vec~\cite{AdiShirsath2021}, emojional~\cite{barry2021emojional}, and JEED1488~\cite{reelfs2020word}.

\subsubsection{Dynamic Embeddings from LLMs}
\label{sec:dynamic-models}
In contrast to static models, Large Language Models (LLMs) generate dynamic, context-aware embeddings. We adopt normalized short names for LLMs throughout (e.g., \textit{gemma-2-2b-it} as \gemmaTwoTwoB, \textit{Llama-3.2-1B-Instruct} as \llamaThreeTwoOneB, \textit{Mistral-7B-Instruct-v0.3} as \mistralVZeroThreeSevenB, and \textit{Qwen2-1.5B-Instruct} as \qwenTwoSevenB) for consistency.

The fundamental principle of these models is their Transformer-based architecture, trained on vast text corpora to understand complex linguistic patterns~\cite{vaswani2017attention}. A key component is their subword \textbf{tokenizer} (e.g., BPE~\cite{sennrich2015neural} or SentencePiece~\cite{kudo2018sentencepiece}), which breaks down text and emojis into a sequence of smaller, known tokens. This mechanism allows LLMs to represent any emoji, even those not seen during training, by composing it from subword units. This provides the comprehensive emoji support that static models lack.

To create an embedding for a sequence (such as an emoji composed of multiple tokens), the LLM processes the entire token sequence. The resulting embedding is \textbf{dynamic} because it is sensitive to context; for example, the representation of an emoji can change depending on the sentence it appears in. For our analysis, which requires context-free, comparable embeddings, we process each emoji as a standalone sequence. To align with our subsequent similarity analyses, we then extract two distinct types of representations from the model's final layer:

\begin{itemize}
    \item \textbf{Aggregated Representation:} We extract the hidden state of the \textbf{final token}~\cite{radford2018improving}. In auto-regressive models, this single vector serves as a holistic embedding by effectively aggregating information from all preceding tokens in the sequence. This representation is used for vector-level metrics like Cosine Distance.
    \item \textbf{Discrete Representation:} We utilize the full sequence of hidden states corresponding to tokens that compose the emoji. This set of vectors preserves token-level detail and supports similarity metrics like Word Mover’s Distance.
\end{itemize}

\subsection{Semantic Similarity Measurement}
\label{sec:similarity_metrics}
To measure semantic similarity, we employ two distinct metrics tailored to different embedding granularities: Cosine Distance for single-vector comparisons and Word Mover’s Distance for sequence-level comparisons.

\subsubsection{Vector-Level Similarity: Cosine Distance}
\label{sec:cos-similarity}
For comparing single-vector representations, such as those from static models or the aggregated output of an LLM, we use Cosine Distance (i.e., $1-$cosine similarity). It quantifies angular dissimilarity and is independent of vector magnitudes. Given two embedding vectors, $V_a$ and $V_b$, the distance is computed as:
\begin{equation}
\text{Cosine Distance}(V_a, V_b) = 1 - \frac{V_a \cdot V_b}{\| V_a \|_2 \times \| V_b \|_2}
\end{equation}
\noindent where $V_a \cdot V_b$ is the dot product and $\|\cdot\|_2$ denotes the Euclidean norm. Lower values indicate higher semantic similarity (more aligned vectors). The vectors can represent either emojis or words, enabling cross-domain comparisons.

\subsubsection{Sequence-Level Similarity: Word Mover’s Distance}
\label{sec:wmd-similarity}
Since LLMs tokenize emojis into sequences of varying lengths, a single-vector metric like Cosine Distance is insufficient. To precisely measure the distance between these token sequences, we adapt the Word Mover’s Distance (WMD)~\cite{kusner2015word}. WMD calculates the minimum cumulative cost required to ``move'' the token embeddings of one sequence to match another.

Given tokenized terms $T_A,T_B$ with embeddings $E_A=\{e_i\}_{i=1}^m\subset\mathbb{R}^d$ and $E_B=\{e'_j\}_{j=1}^n\subset\mathbb{R}^d$, the WMD is
\begin{equation}
\mathrm{WMD}(T_A,T_B)=\min_{F\in\mathbb{R}_+^{m\times n}}\sum_{i=1}^m\sum_{j=1}^n F_{ij}\,c(e_i,e'_j)
\end{equation}
\noindent subject to\ $\sum_{j=1}^n F_{ij}=w_i$ and $\sum_{i=1}^m F_{ij}=w'_j$ with $w_i=\tfrac{1}{m}$, $w'_j=\tfrac{1}{n}$. Where $F_{ij}$ is the transport flow and $c(\cdot,\cdot)$ is the ground distance (\textbf{Euclidean} or \textbf{Cosine}).

\subsection{Bias Measurements}
\label{sec:bias-measurement}

\textbf{RND~\cite{garg2018word}:} Relative Norm Distance (RND) is a metric that quantifies bias by measuring whether a set of neutral words is, on average, closer in an embedding space to one target group of words than to another. In our study, we use RND to assess the bias between two groups of skin-toned emojis (e.g., light-toned vs. dark-toned) relative to a set of neutral words. This neutral word set is derived from the NRC-VAD lexicon~\cite{mohammad2018obtaining}, selected based on a neutral valence score range. The RND score is computed by summing the difference in Euclidean distance from each neutral word vector ($V_m$) to the centroid of the first group ($V_1$) and the centroid of the second group ($V_2$):
\begin{equation}
\text{RND Score} = \sum_{V_m \in M} (\| V_m - V_1 \|_2 - \| V_m - V_2 \|_2)
\end{equation}
\noindent A positive score indicates that the neutral words are collectively closer to the centroid of the second group ($V_2$), while a negative score suggests a closer association with the centroid of the first group ($V_1$).

\textbf{WEAT~\cite{caliskan2017semantics}:} The Word Embedding Association Test (WEAT) is a statistical method for measuring implicit associations between concepts in embedding spaces. In our study, we use it to quantify potential biases by measuring the association between sets of \textit{target} emojis (e.g., different skin tones) and sets of \textit{attribute} words (e.g., positive vs. negative concepts).

The test is framed around a null hypothesis, $H_0$, which posits that there is no difference in the relative association of the two target sets with the two attribute sets. The statistical significance (p-value) against this hypothesis is determined via a permutation test, which assesses the probability of observing the measured association by random chance.

Given two target sets $X$ and $Y$ and two attribute sets $A$ and $B$, the WEAT test statistic $s(X,Y,A,B)$ is calculated as:
\begin{equation}
s(X,Y,A,B) = \sum_{x \in X} s(x,A,B) - \sum_{y \in Y} s(y,A,B)
\end{equation}
\noindent where $s(w,A,B)$ measures the differential association of a single target item $w$ with the attributes:
\begin{equation}
s(w,A,B) = \text{mean}_{a \in A} \cos(\vec{w},\vec{a}) - \text{mean}_{b \in B} \cos(\vec{w},\vec{b})
\end{equation}
A large positive value of $s(X,Y,A,B)$ indicates a stronger association of set $X$ with set $A$ (and $Y$ with $B$). The magnitude of this association is measured by the effect size:
\begin{equation}
\text{Effect Size} = \frac{\text{mean}_{x \in X} s(x,A,B) - \text{mean}_{y \in Y} s(y,A,B)}{\text{std\_dev}_{w \in X \cup Y} s(w,A,B)}
\end{equation}

\textbf{RNSB~\cite{sweeney2019transparent}:} Relative Negative Sentiment Bias (RNSB) is a metric designed to quantify how disproportionately a set of terms is associated with negative sentiment. The methodology first involves training a logistic regression classifier on a gold-standard set of word embeddings with known positive and negative sentiment labels ($S$). This process learns a set of weights ($\theta$) that define a ``negative sentiment'' direction in the embedding space by minimizing a regularized logistic loss function:
\[
\min_{\theta \in \mathbb{R}^d} \sum_{(x_i, y_i) \in S} l(y_i, \theta^T x_i) + \lambda \|\theta\|^2
\]
Once this classifier ($f^*$) is trained, it is used to predict the negative sentiment probability for each vector in a target set of $t$ emojis, $K = \{k_1, ..., k_t\}$. These individual probabilities are subsequently normalized to form a probability distribution $P$ over the set:
\[
P = \left\{\frac{f^*(k_1)}{\sum_{j=1}^{t}f^*(k_j)},...,\frac{f^*(k_t)}{\sum_{j=1}^{t}f^*(k_j)}\right\}
\]
The final RNSB score is then calculated as the Kullback-Leibler (KL) divergence between this observed distribution $P$ and a uniform distribution $U$, where each emoji would have a $1/t$ probability:
\[
RNSB(P) = D_{KL}(P||U)
\]
A higher RNSB score indicates a greater deviation from fairness, signifying that negative sentiment is unevenly distributed across the target set of emojis.

\section{Model Coverage and Tokenization of Skin‑Tone Emojis}
\label{sec:exp_models} 

\subsection{Representational Gaps in Static Embedding}
\label{sec:exp-static-emoji}
We benchmark a range of widely-used static embedding models to evaluate their support for skin-toned emojis, with detailed results in Table~\ref{tab:list_of_pretrained_models}. Our evaluation revealed a significant gap in coverage, as the majority of the nine surveyed models offer minimal support for skin tone modifiers. In contrast, \textbf{emoji2vec}~\cite{eisner2016emoji2vec} and \textbf{emoji-sw2v}~\cite{barbieri2018gender} provide substantially better coverage. Consequently, while our initial benchmark is broad, our subsequent semantic and bias analyses will concentrate on these two primary models, treating the others as important but secondary points of comparison for the state of static emoji embedding support.

\begin{table}[ht] 
    \setlength{\tabcolsep}{4pt}
    \centering 
    \caption{Support for skin-toned emojis in pre-trained embedding models. The table lists the total vocabulary size (\# Tokens) for each model and the number of supported skin-toned emojis, with the total number of supported emojis shown in parentheses.}~\label{tab:list_of_pretrained_models}
    \begin{tabular}{llll}
        \toprule
        \textbf{Pre-trained Model} & \textbf{\# Tokens} & \textbf{\# Skin-toned Emojis} \\
        \midrule
        FastText-crawl-300d-2M~\cite{bojanowski2016enriching} & 1,999,995 & 5 (642) \\ 
        GloVe-Twitter-27B~\cite{pennington2014glove} & 1,193,514 & 0 (139) \\
        FastText-wiki-news~\cite{bojanowski2016enriching} & 999,994 & 1 (147) \\ 
        emoji-word2vec~\cite{AdiShirsath2021} & 23,069 & 5 (930) \\ 
        emoji-sw2v~\citep{barbieri2018gender} & 2269 & 834 (2269) \\ 
        emojional~\cite{barry2021emojional} & 1816 & 0 (1816) \\ 
        emoji2vec~\cite{eisner2016emoji2vec} & 1661 & 280 (1661) \\ 
        JEED1488~\cite{reelfs2020word} & 1488 & 0 (1488) \\ 
        BERT~\cite{devlin2018bert} & 30,522 & 0 (0) \\
        \bottomrule
    \end{tabular}
\end{table}

\subsection{Representation and Tokenization in LLMs}
\label{sec:exp-llm-emoji}

A key advantage of modern LLMs is that their tokenizers are inherently equipped to handle the full diversity of skin-toned emojis, ensuring representational fairness at a foundational level. Unlike static models constrained by fixed vocabularies, a tokenizer's ability to decompose any character sequence guarantees that all official skin-toned emojis are correctly segmented and represented. While this comprehensive coverage is a universal feature, the tokenization strategies and resulting efficiency differ across models. To quantify this, we analyzed the tokenization statistics for all 2,735 skin-toned emojis across several model families: \textbf{\llamaThreeTwoThreeB}~\cite{meta_llama3.2_3b_instruct}, \textbf{\qwenTwoSevenB}~\cite{qwen2}, \textbf{\gemmaTwoNineB}~\cite{gemma_2024}, and \textbf{\mistralVZeroThreeSevenB}~\cite{mistral7b_instruct_v0.3}. The summary statistics are presented in ~\Cref{tab:tokenizer_stats}, and a more detailed visualization is available in Appendix~\ref{sec:appendix_token_dist}.

To investigate the source of these differences, we isolated the skin tone modifiers themselves and analyzed the number of tokens each model requires to represent them, as shown in Table~\ref{tab:modifier_token_length}. The results reveal a critical disparity: Gemma and Qwen exhibit perfect consistency, encoding each modifier as a single token. Llama is also consistent, using three tokens for each. Mistral, however, shows a significant bias; it requires more tokens (2) for modifiers than the most efficient models and, strikingly, demands a substantially higher token count (5) specifically for the dark skin tone modifier. This inconsistency suggests a form of computational bias, where representing a specific demographic attribute incurs a disproportionately higher processing overhead, indicating a \textit{foundational inequity in the model's architecture}. Practically, this inflation in token count translates to increased computational latency and API costs, subtly penalizing the processing of diverse identities in the web.

\begin{table}[ht] 
    \centering 
    \caption{Tokenization statistics for skin-toned emojis across LLMs. Columns show each model's average tokens per emoji (Avg. Tokens/Emoji), minimum to maximum token range (Token Range), and the most frequent token count (mode), with the number of emojis at that count in parentheses.}~\label{tab:tokenizer_stats}
    \resizebox{\linewidth}{!}{
    \begin{tabular}{lrrr}
        \toprule
        \textbf{Model} & \makecell{\textbf{Avg. Tokens/}\\\textbf{Emoji}} & \makecell{\textbf{Token}\\\textbf{Range}} & \makecell{\textbf{Most Common}\\\textbf{Count}} \\
        \midrule
        \gemmaTwoTwoB~\cite{gemma_2024} & 3.96 & 1--9 & 3 (893 emojis) \\ 
        \qwenTwoSevenB~\cite{qwen2} & 5.97 & 1--13 & 6 (635 emojis) \\
        Llama-3.2-1B/3B~\cite{meta_llama3.2_1b_instruct,meta_llama3.2_3b_instruct} & 10.05 & 3--19 & 10 (550 emojis) \\ 
        \mistralVZeroThreeSevenB~\cite{mistral7b_instruct_v0.3} & $\sim$9.5 & 2--26 & 6 (490 emojis) \\ 
        \bottomrule
    \end{tabular}
    }
\end{table}

\begin{table}[h]
\centering
\caption{Number of tokens for each of the five skin tone modifiers across different LLMs. The inconsistency in Mistral's tokenization for the dark skin tone is highlighted.}
\label{tab:modifier_token_length}
\begin{tabular}{lccccc}
\toprule
\textbf{Model} & \includegraphics[width=0.04\linewidth]{figures/emoji_image/light.jpg} & \includegraphics[width=0.04\linewidth]{figures/emoji_image/med_light.jpg} & \includegraphics[width=0.04\linewidth]{figures/emoji_image/med.jpg} & \includegraphics[width=0.04\linewidth]{figures/emoji_image/med_dark.jpg} & \includegraphics[width=0.04\linewidth]{figures/emoji_image/dark.jpg} \\
\midrule
Gemma-2 & 1 & 1 & 1 & 1 & 1 \\
Qwen-2 & 1 & 1 & 1 & 1 & 1 \\
Llama-3 & 3 & 3 & 3 & 3 & 3 \\
Mistral-v0.7 & 2 & 2 & 2 & 2 & \textcolor{red}{\textbf{5}} \\
\bottomrule
\end{tabular}
\end{table}

\section{Semantic Impact of Skin Tone Modifiers}
\label{sec:exp-semantic-consist}
To comprehensively evaluate the semantic impact of skin tone modifiers, we conducted a three-part analysis investigating potential semantic drift from distinct but complementary perspectives. We assess whether skin tone variations affect: the semantic consistency between an emoji and its textual description; the internal semantic consistency among variants of hand emoji families; and the alignment with predefined affective polarity (i.e., `Good' vs. `Bad').

\subsection{Emoji–Text Semantic Alignment}
\label{sec:exp-emoji-text-align}
An emoji and its textual description (e.g., \includegraphics[width=0.04\linewidth]{figures/emoji_image/U+1F44C.png} and ``OK hand'') should ideally be semantically equivalent. Given that LLMs can process both modalities, this section investigates whether skin tone modifiers introduce semantic drift, causing a discrepancy between an emoji's representation and that of its corresponding text. To explore this, we analyze semantic consistency across our suite of LLMs using two representation strategies outlined in~\Cref{sec:dynamic-models}.

First, for the \textbf{Aggregated Representation}, we measure the Cosine Distance between the final hidden state vector of an emoji and that of its textual description. Second, for the \textbf{Discrete Representation}, we compute the WMD between the full sequence of token embeddings for the emoji and its text, using both Euclidean and Cosine ground distances. A smaller distance indicates better semantic alignment. By comparing these metrics across the six skin tone categories (Default and the five modifiers), we can identify whether certain skin tones consistently cause a greater semantic deviation, indicating a representational bias. The mean results for each model are summarized in~\Cref{tab:semantic_consistency}.

\begin{table}[ht]
\centering
\caption{Semantic consistency between emojis and their textual descriptions across different LLMs and skin tones. All cell values represent the \textbf{mean} score for the given metric. For Cosine Distance and WMD, higher values indicate more semantic shift between the cross-modal representations.}
\label{tab:semantic_consistency}
\resizebox{\linewidth}{!}{
\begin{tabular}{lcccccc}
\toprule
\textbf{Model} & \textbf{Default} & \includegraphics[width=0.04\linewidth]{figures/emoji_image/light.jpg} & \includegraphics[width=0.04\linewidth]{figures/emoji_image/med_light.jpg} & \includegraphics[width=0.04\linewidth]{figures/emoji_image/med.jpg} & \includegraphics[width=0.04\linewidth]{figures/emoji_image/med_dark.jpg} & \includegraphics[width=0.04\linewidth]{figures/emoji_image/dark.jpg} \\
\midrule
\multicolumn{7}{c}{\textbf{Aggregated Representation Distance} (Cosine Distance)} \\
\midrule
\gemmaTwoTwoB & 0.819 & 0.906 & 0.864 & 0.860 & 0.856 & \textbf{0.912} \\
\gemmaTwoNineB & 0.612 & 0.718 & \textbf{0.794} & 0.746 & 0.778 & 0.695 \\
\llamaThreeTwoOneB & \textbf{0.740} & 0.662 & 0.653 & 0.668 & 0.641 & 0.694 \\
\llamaThreeTwoThreeB & \textbf{0.721} & 0.644 & 0.661 & 0.643 & 0.651 & 0.639 \\
\llamaThreeOneEightB & \textbf{0.701} & 0.568 & 0.577 & 0.570 & 0.588 & 0.578 \\
\mistralVZeroThreeSevenB & \textbf{0.797} & 0.737 & 0.709 & 0.733 & 0.722 & 0.728 \\
\qwenTwoSevenB & \textbf{1.001} & 0.896 & 0.844 & 0.824 & 0.775 & 0.786 \\
\midrule
\multicolumn{7}{c}{\textbf{Discrete Representation Distance} (WMD-Cosine)} \\
\midrule
\gemmaTwoTwoB & 0.865 & 0.906 & 0.936 & 0.885 & 0.922 & 0.883 \\
\gemmaTwoNineB & 0.806 & 0.866 & 0.885 & 0.853 & 0.882 & 0.851 \\
\llamaThreeTwoOneB & 0.944 & 0.947 & 0.935 & 0.942 & 0.933 & 0.939 \\
\llamaThreeTwoThreeB & 0.942 & 0.950 & 0.934 & 0.940 & 0.930 & 0.941 \\
\llamaThreeOneEightB & 0.965 & 0.966 & 0.955 & 0.956 & 0.955 & 0.956 \\
\mistralVZeroThreeSevenB & 0.968 & 0.966 & 0.925 & 0.955 & 0.922 & 0.953 \\
\qwenTwoSevenB & 0.946 & 0.947 & 0.931 & 0.936 & 0.938 & 0.934 \\
\midrule
\multicolumn{7}{c}{\textbf{Discrete Representation Distance} (WMD-Euclidean)} \\
\midrule
\gemmaTwoTwoB & 2.281 & 2.493 & 2.517 & 2.444 & 2.565 & 2.432 \\
\gemmaTwoNineB & 2.096 & 2.254 & 2.245 & 2.198 & 2.272 & 2.190 \\
\llamaThreeTwoOneB & 1.334 & 1.353 & 1.331 & 1.341 & 1.329 & 1.330 \\
\llamaThreeTwoThreeB & 1.503 & 1.534 & 1.505 & 1.522 & 1.503 & 1.516 \\
\llamaThreeOneEightB & 0.837 & 0.830 & 0.824 & 0.823 & 0.833 & 0.825 \\
\mistralVZeroThreeSevenB & 0.220 & 0.225 & 0.216 & 0.223 & 0.216 & 0.222 \\
\qwenTwoSevenB & 1.019 & 1.039 & 1.011 & 1.010 & 1.002 & 0.994 \\
\bottomrule
\end{tabular}
}
\end{table}

The results in~\Cref{tab:semantic_consistency}, particularly the cosine distance measurements, reveal several key patterns. First, model scale appears to influence performance; for both Gemma and Llama, the largest models (9B and 8B, respectively) achieve significantly better semantic alignment than their smaller counterparts. We also observe distinct model-specific biases. For instance, the Gemma model shows a potential semantic drift, where the alignment for dark skin tones is noticeably weaker than for light skin tones. Conversely, Qwen exhibits a different and consistent trend where the alignment between an emoji and its text improves as skin tone becomes darker.

\subsection{Within‑Emoji Variant Consistency}
\label{sec:exp-tsne}
To intuitively understand the semantic relationships between skin tone variants, we conducted a t-SNE visualization analysis. This experiment examines the relationships between a default emoji and its different skin tone variants within the semantic space. We specifically investigate whether the variants form a tight cluster based on their core semantic meaning or separate by skin tone, and if they drift apart due to the skin tone modifier.

\begin{figure*}[ht]
    \centering
    \includegraphics[width=0.15\linewidth]{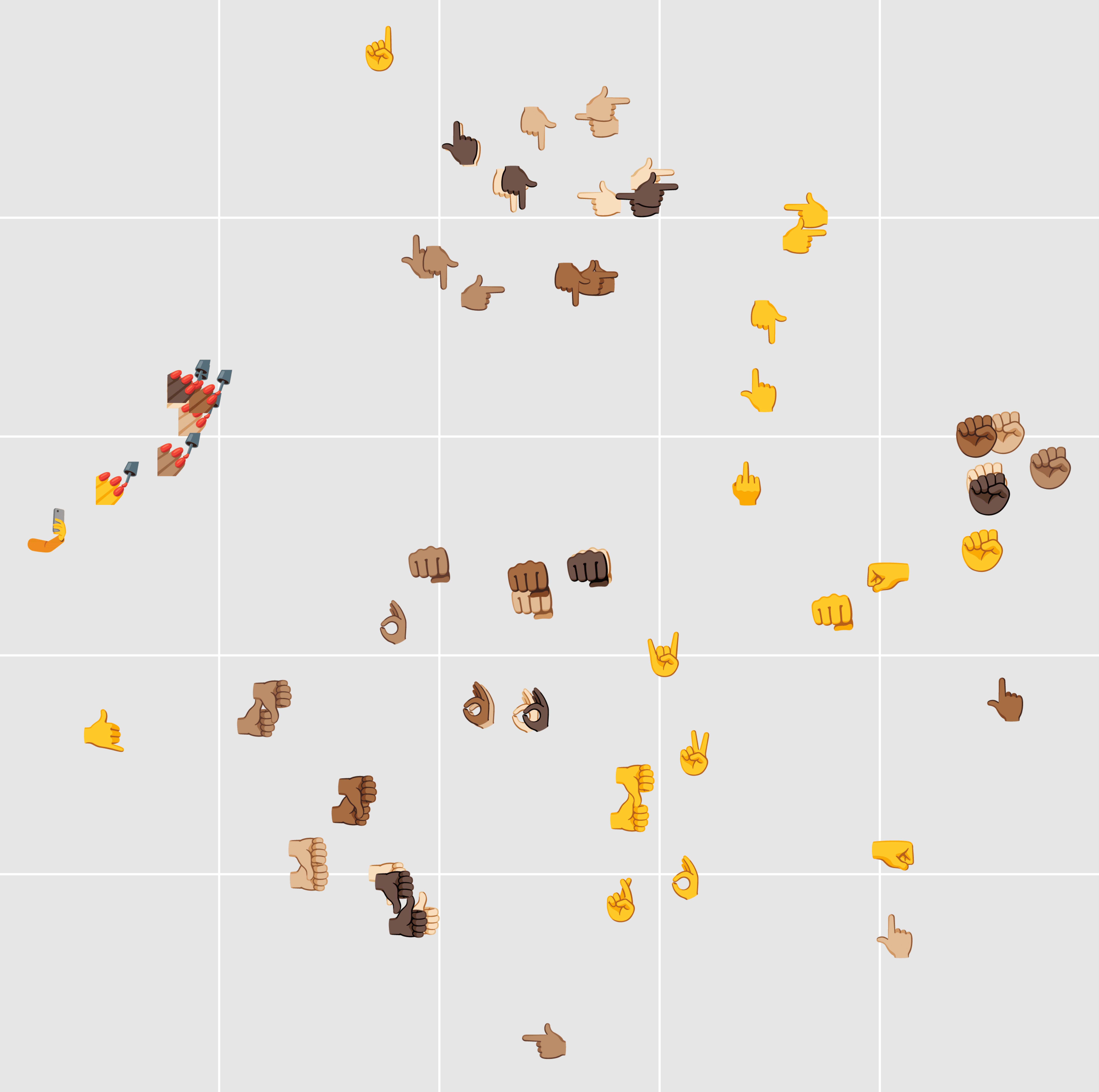}\hfill
    \includegraphics[width=0.15\linewidth]{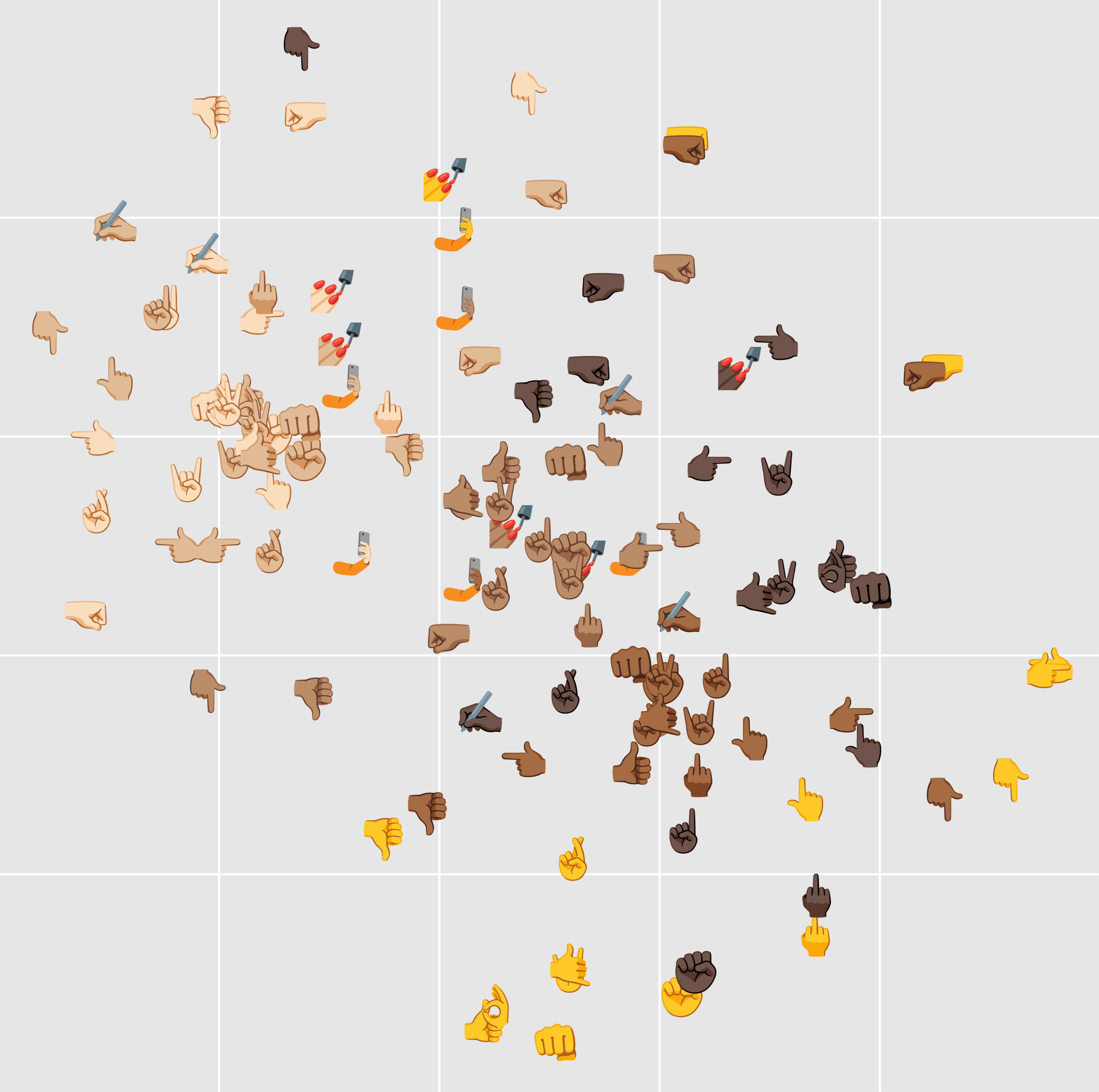}\hfill
    \includegraphics[width=0.15\linewidth]{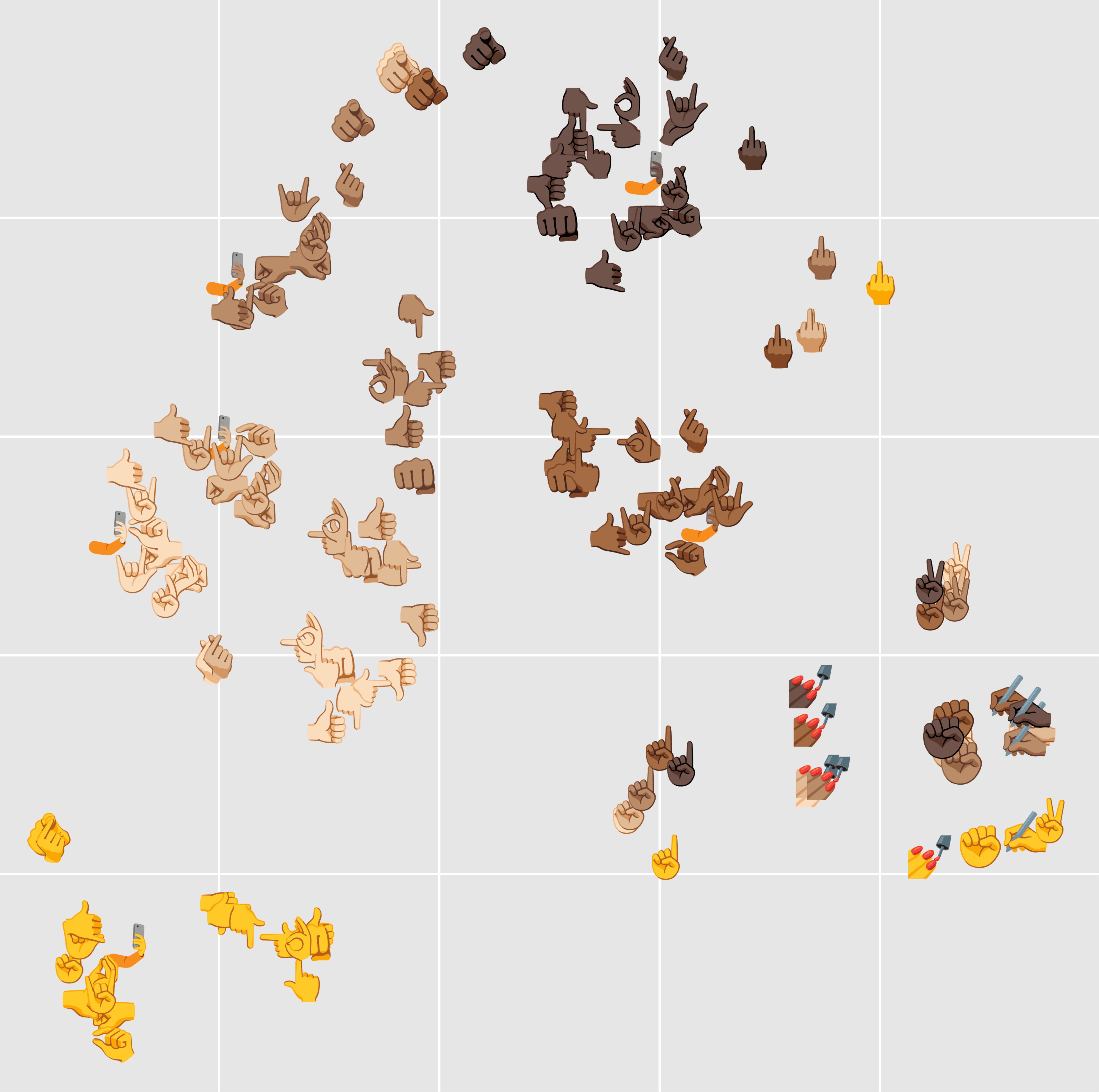}\hfill
    \includegraphics[width=0.15\linewidth]{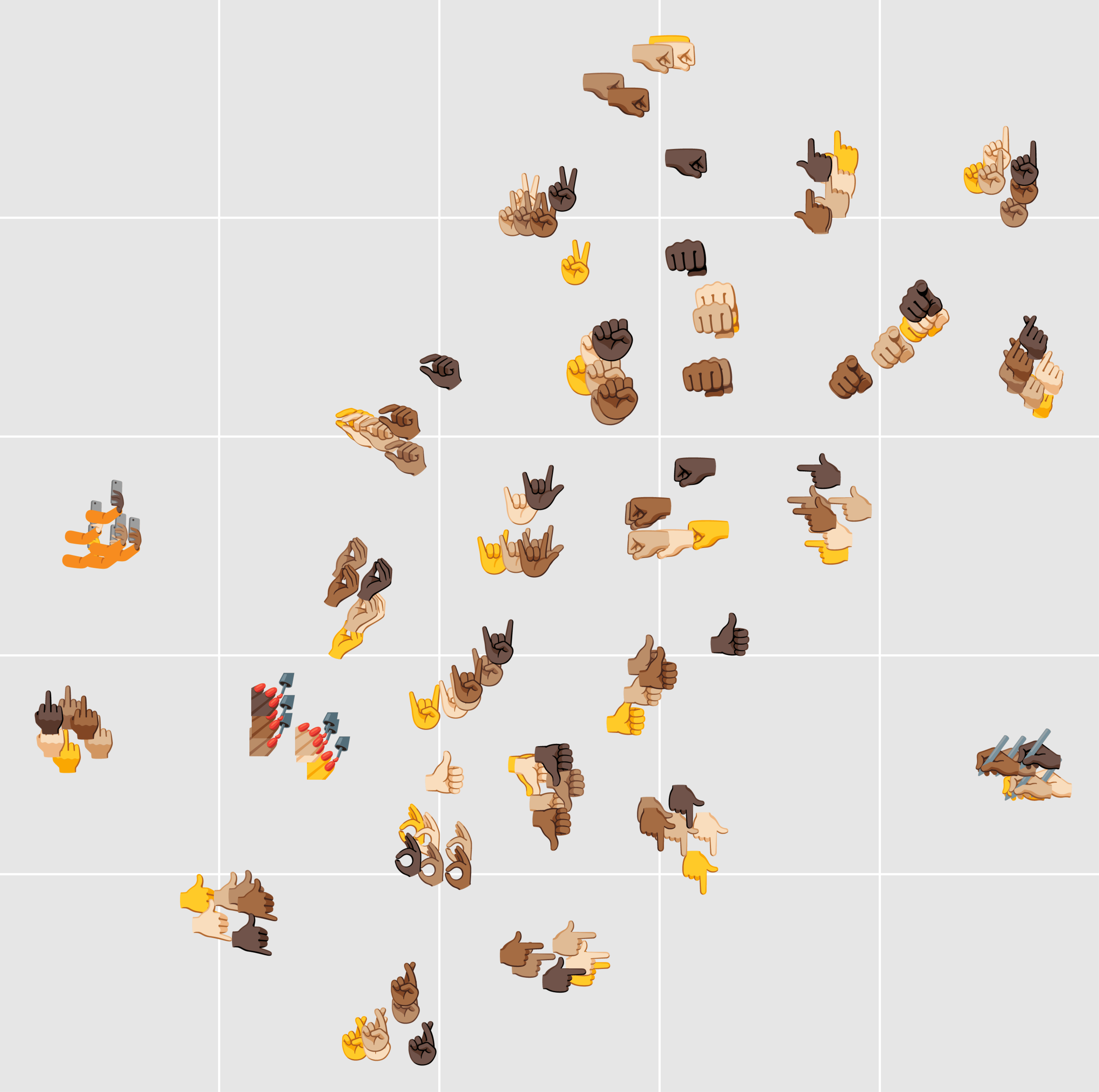}\hfill
    \includegraphics[width=0.15\linewidth]{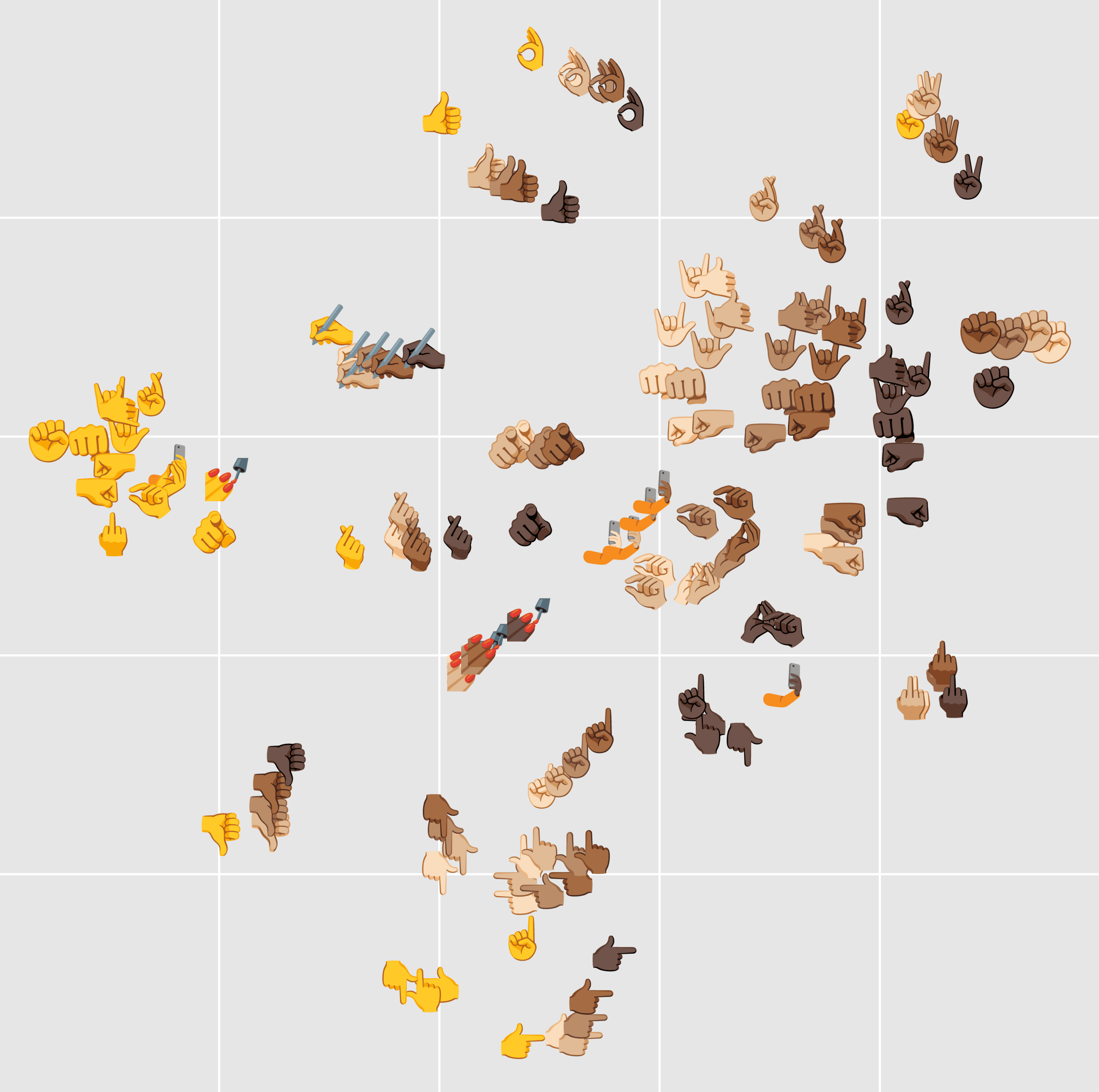}\hfill
    \includegraphics[width=0.15\linewidth]{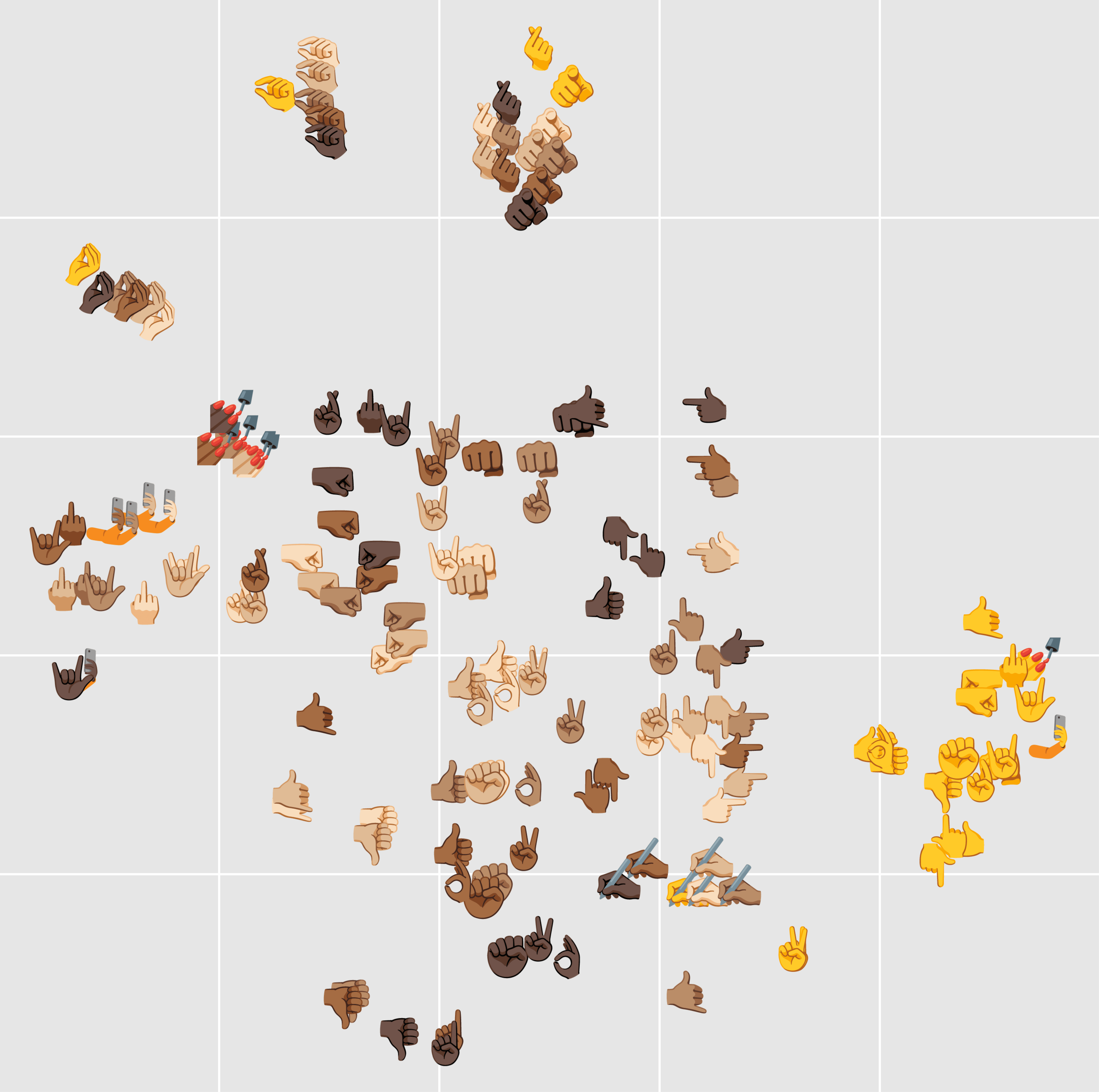}\\[0.5em]

    \subfigure[Emoji2vec]{
        \includegraphics[width=0.15\linewidth]{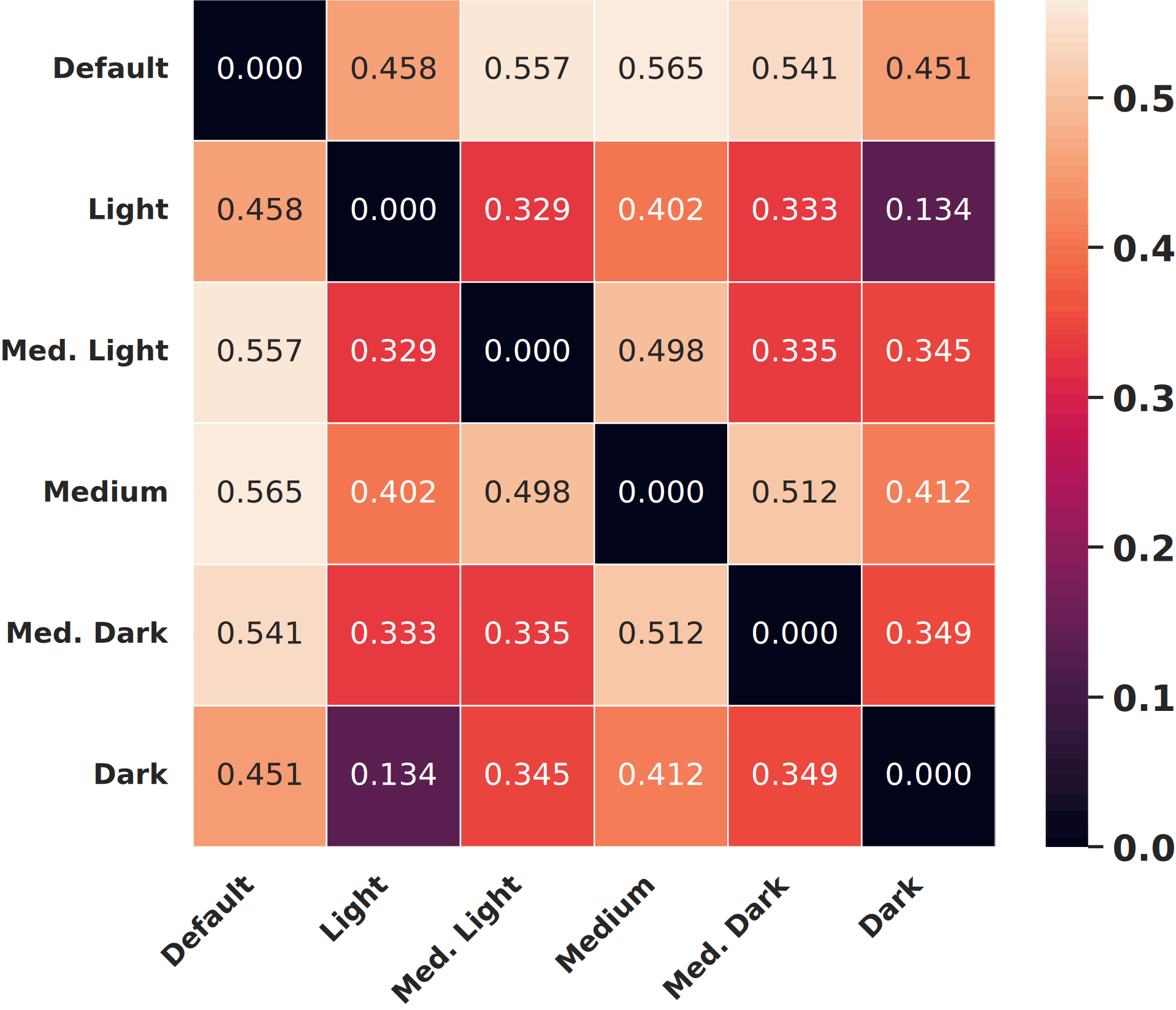}
        \label{fig:heatmap-emoji2vec}
    }\hfill
    \subfigure[SW2V]{
        \includegraphics[width=0.15\linewidth]{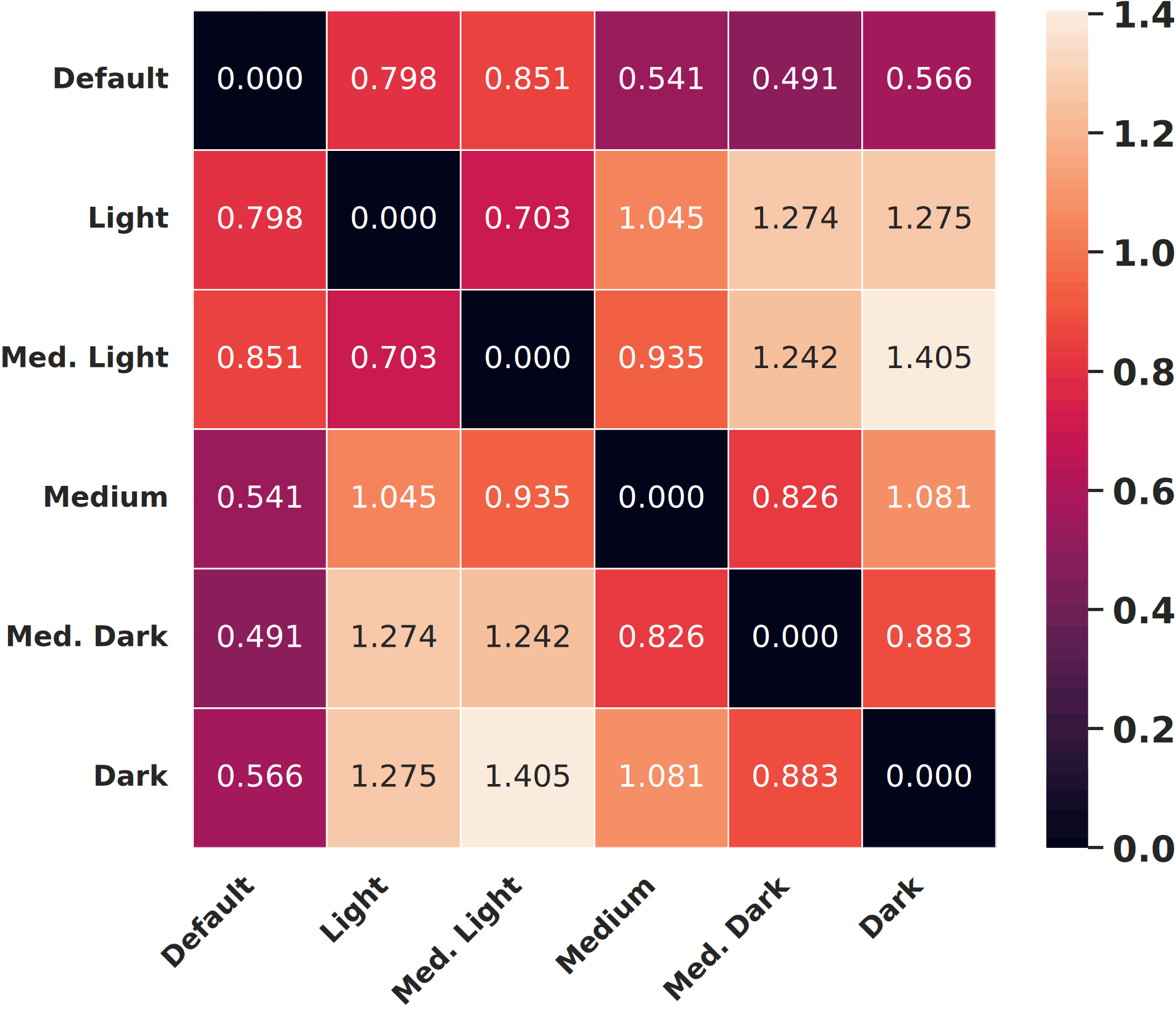}
        \label{fig:heatmap-sw2v}
    }\hfill
    \subfigure[\llamaThreeTwoOneB]{
        \includegraphics[width=0.15\linewidth]{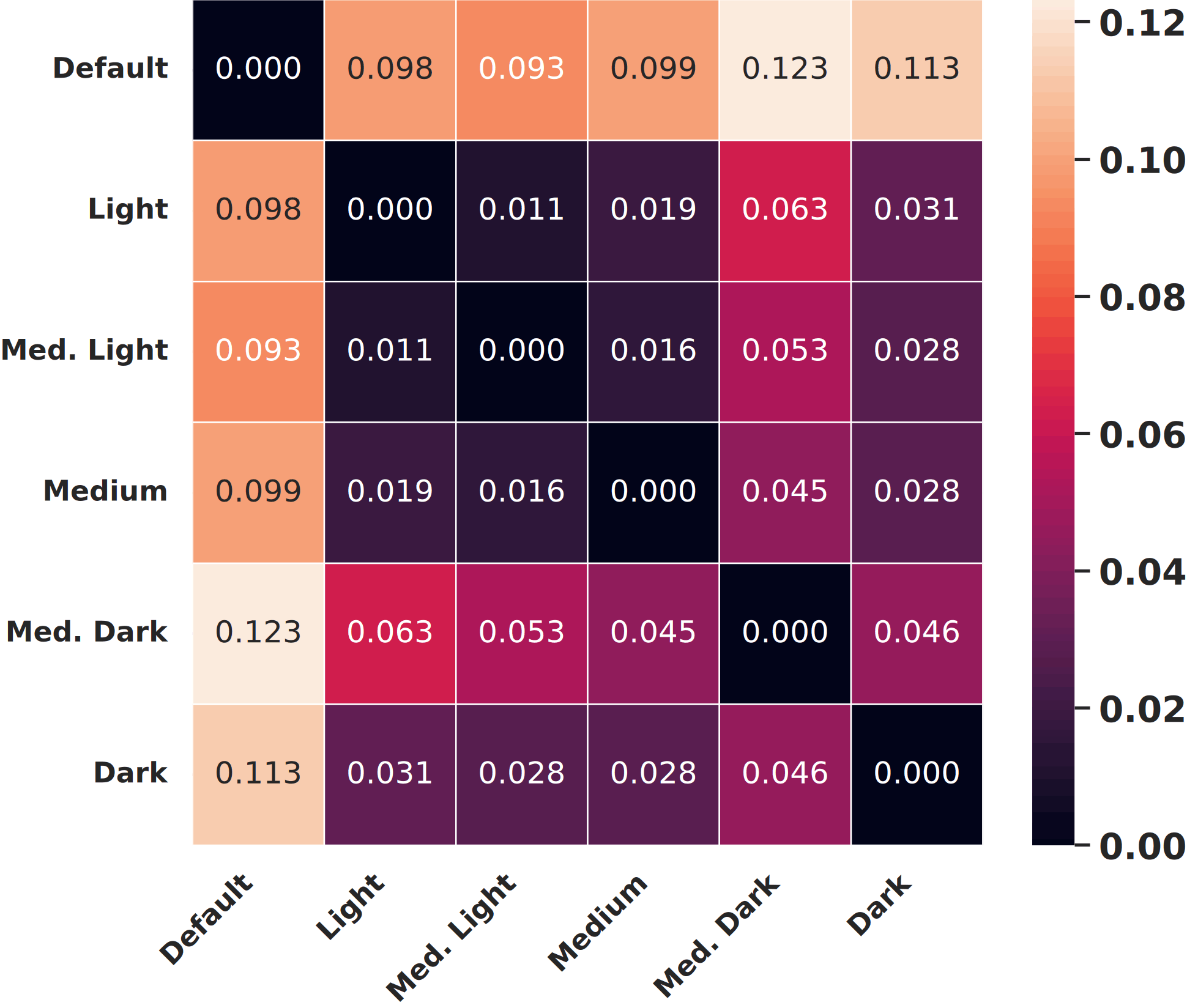}
        \label{fig:heatmap-llama-1b}
    }\hfill
    \subfigure[\gemmaTwoTwoB]{
        \includegraphics[width=0.15\linewidth]{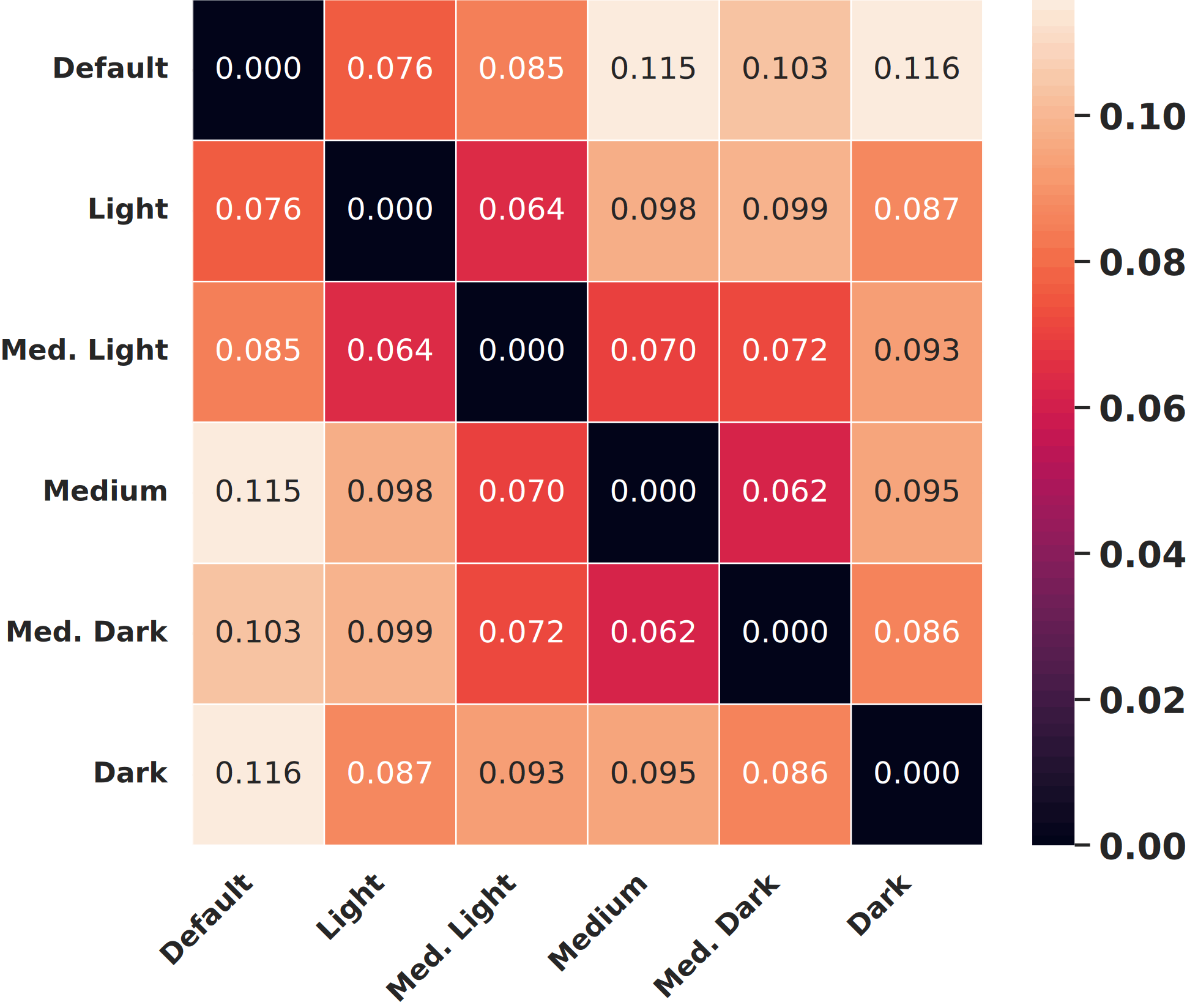}
        \label{fig:heatmap-gemma-2b}
    }\hfill
    \subfigure[\mistralVZeroThreeSevenB]{
        \includegraphics[width=0.15\linewidth]{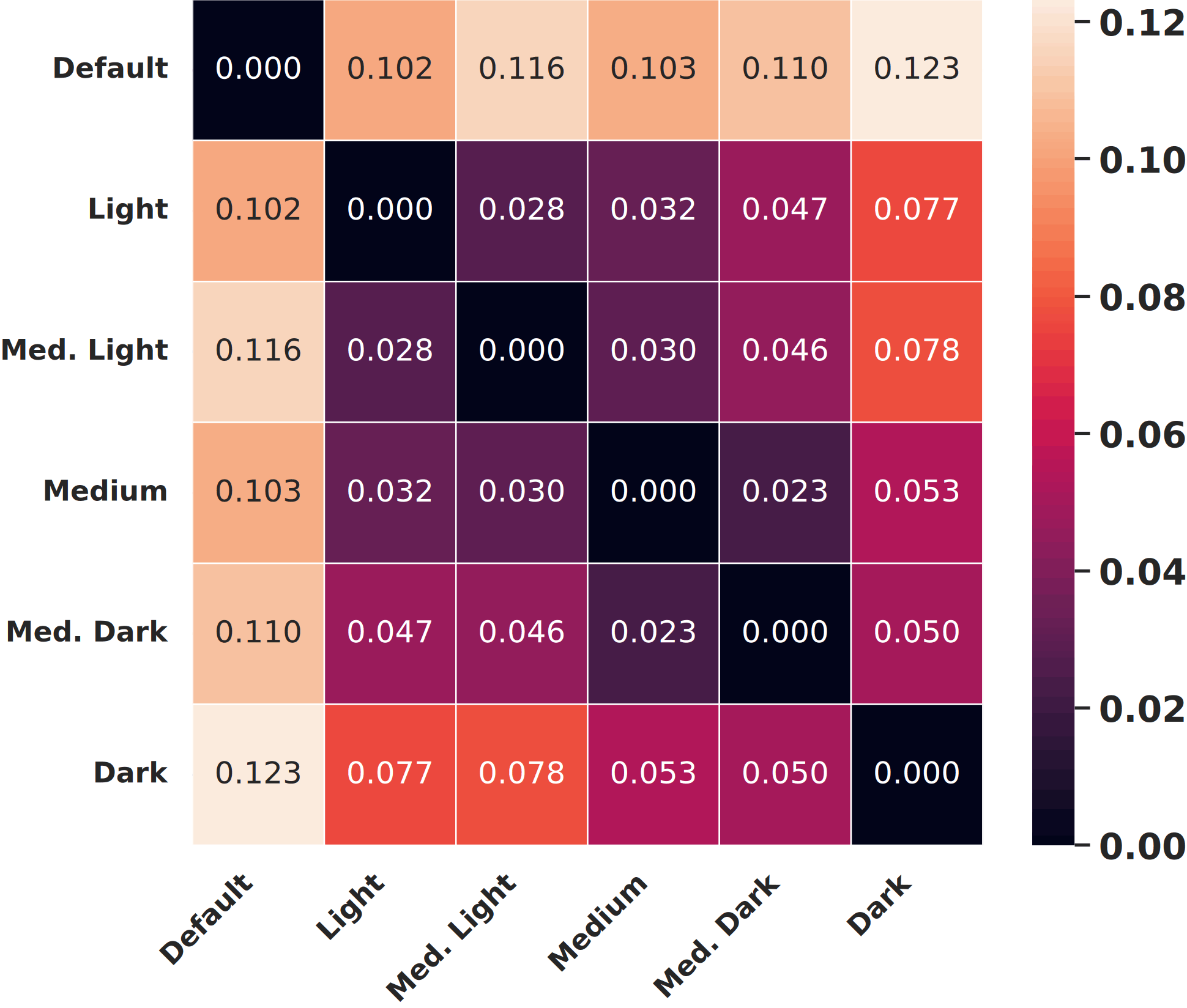}
        \label{fig:heatmap-mistral-7b}
    }\hfill
    \subfigure[\qwenTwoSevenB]{
        \includegraphics[width=0.15\linewidth]{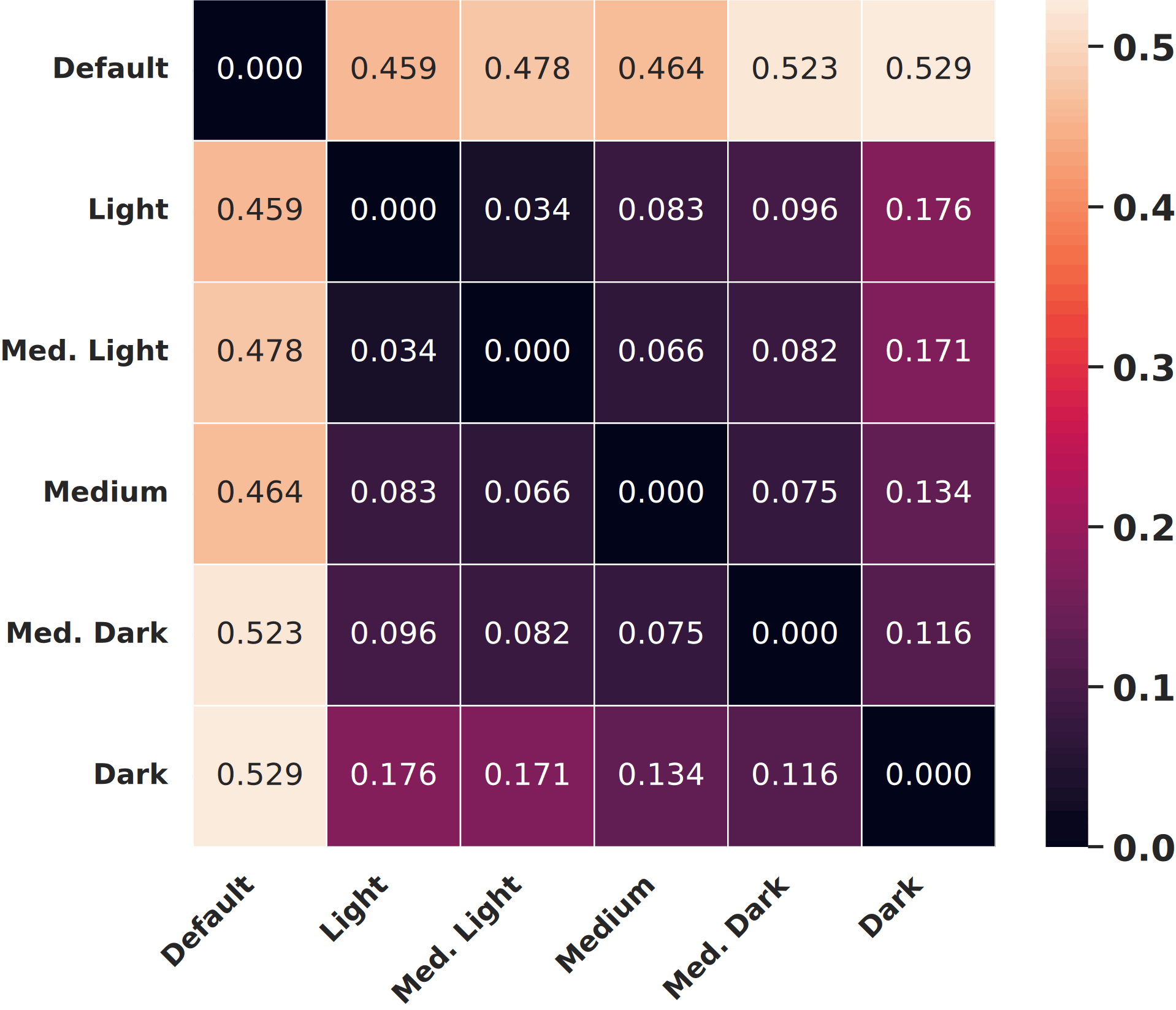}
        \label{fig:heatmap-qwen-1.5b}
    }
    \vspace{-0.5em}
    \caption{t-SNE visualizations (top row) and similarity shift heatmaps (bottom row) for \textit{hand gesture} emoji embeddings across six models. Columns correspond to the same model as the caption shown. The t-SNE plots qualitatively illustrate clustering patterns of emoji variants, while the heatmaps depict pairwise \textbf{cosine-distance} shifts between skin-tone variants (higher values indicate larger semantic shifts).}
    \label{fig:tsne_heatmap_combined}
\end{figure*}

We deliberately selected \textit{hand gesture} emojis from the Unicode standard test set\footnote{https://www.unicode.org/reports/tr51}. This category was chosen for its rich semantic diversity, as hand gestures often carry dual meanings, conveying both emotional states (e.g., \includegraphics[width=0.04\linewidth]{figures/emoji_image/U+1F44C.png} is fine) and functional descriptions (e.g., \includegraphics[width=0.04\linewidth]{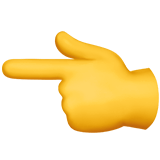} is pointing left). We generated embeddings for this emoji set using our six primary models: the two static models (emoji2vec and emoji-sw2v) and the four LLMs (\llamaThreeTwoOneB, \gemmaTwoTwoB, \mistralVZeroThreeSevenB, and \qwenTwoSevenB), using their aggregated representations. These high-dimensional vectors were then projected into a 2D space using the t-SNE algorithm. The resulting visualizations, presented in~\Cref{fig:tsne_heatmap_combined}, allow for a qualitative comparison of the clustering patterns across the different model architectures, revealing how each model groups or separates emojis based on their core meaning versus their skin tone.

\Cref{fig:tsne_heatmap_combined} reveals distinct clustering strategies and biases across the models. The static models diverge significantly: emoji2vec primarily groups emojis by their core semantic meaning, but most skin tone variants exhibit a large shift from the default emoji. In contrast, emoji-sw2v demonstrates a gradient clustering pattern based on skin tone, where the light skin tone variants are systematically positioned furthest from the default.

Among the LLMs, each model adopts a different clustering approach. Gemma exhibits the highest degree of fairness, organizing variants by semantics with all skin tones tightly clustered around the default emoji. Conversely, Llama serves as a clear example of skin-tone-based clustering, where darker skin tones are consistently distanced furthest from the default. Qwen and Mistral display a similar bias, also showing a notable separation between the dark skin tone variants and the default emoji, suggesting a prevalent pattern of representational inequity.

\subsection{Fine‑Grained Pairwise Tone Similarity}
\label{sec:exp-heatmap}
To further probe representational differences across skin tones, we conduct a cross–skin-tone analysis using similarity heatmaps. Following the representation procedure in~\Cref{sec:dynamic-models} and the similarity definition in~\Cref{sec:cos-similarity}, we obtain aggregated embeddings for each emoji variant and compute, for each model, the pairwise \textit{cosine distance} between the same base emoji under two tone variants. Concretely, for every base emoji, we evaluate the Cosine Distance (1$-$cosine similarity; see~\Cref{sec:cos-similarity}) for each distinct tone pair and average across emojis to populate the matrix. Each cell therefore reports the mean cosine distance for a tone pair: higher values indicate larger semantic shifts attributable to the skin tone modifier, whereas lower values suggest the variants are treated as stylistic variations (i.e., lower sensitivity to skin tone). 

\Cref{fig:tsne_heatmap_combined} reveals distinct patterns in how pair-wise emoji skin tones are represented. Notably, Mistral demonstrates a more stable semantic space, exhibiting smaller vector changes between different skin tones. A clear divergence is observed between model families: static embedding models show significantly larger distances between skin tone variations compared to LLMs, which generally maintain closer representations. Furthermore, most LLMs exhibit a systematic semantic drift where the representational distance between lighter and darker skin tones increases, suggesting a shift in embedding space that correlates with the skin tone spectrum.

\section{Bias in Tone‑modified Emoji Embeddings}
\label{sec:exp-bias}
This section presents our primary contribution: a systematic quantification of biases embedded in skin-toned emoji representations. Our analysis includes Relative Norm Distance (RND), the Word Embedding Association Test (WEAT), and Relative Negative Sentiment Bias (RNSB) introduced in~\Cref{sec:similarity_metrics}. We conduct this investigation across five models: one static model (emoji2vec) and four LLMs (\llamaThreeTwoThreeB, \gemmaTwoTwoB, \qwenTwoSevenB, and \mistralVZeroThreeSevenB). For models requiring separate text embeddings, such as emoji2vec, we utilize its base model, the Word2Vec model~\cite{mikolov2013efficient}.

\subsection{Neutral Words Bias via RND}
\label{sec:exp-rnd}
Our first analysis tests whether skin tone modifiers induce a semantic shift relative to a \textit{neutral baseline}. We use Relative Norm Distance (RND; see~\Cref{sec:bias-measurement}) to assess whether a neutral word set is, on average, closer to one skin-tone group than another. The neutral words are drawn from the NRC-VAD lexicon~\cite{mohammad2018obtaining}, filtered to a neutral valence range (0.48--0.52). A non-zero score indicates directional bias (the sign identifies which group is closer, and larger absolute values indicate stronger bias). Consistent with the setup in~\Cref{sec:exp-tsne}, we restrict the analysis to the hand gestures emoji subdomain to isolate the specific impact of the skin tone modifier.

\begin{figure*}[ht]
    \centering
    \subfigure[Emoji2vec]{
        \includegraphics[width=0.18\linewidth]{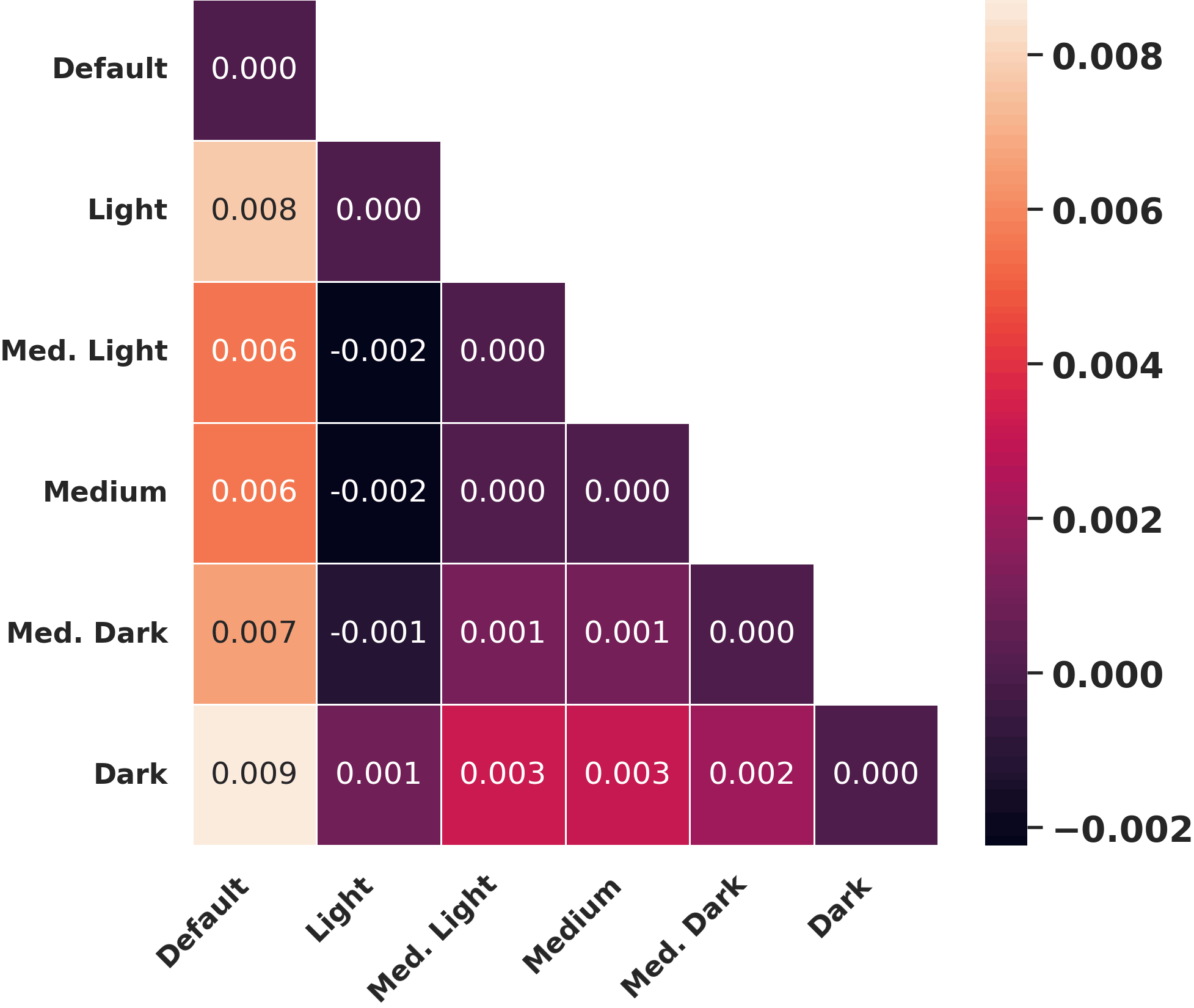}
        \label{fig:rnd_heatmap_e2v}
    }
    \subfigure[\gemmaTwoTwoB]{
        \includegraphics[width=0.18\linewidth]{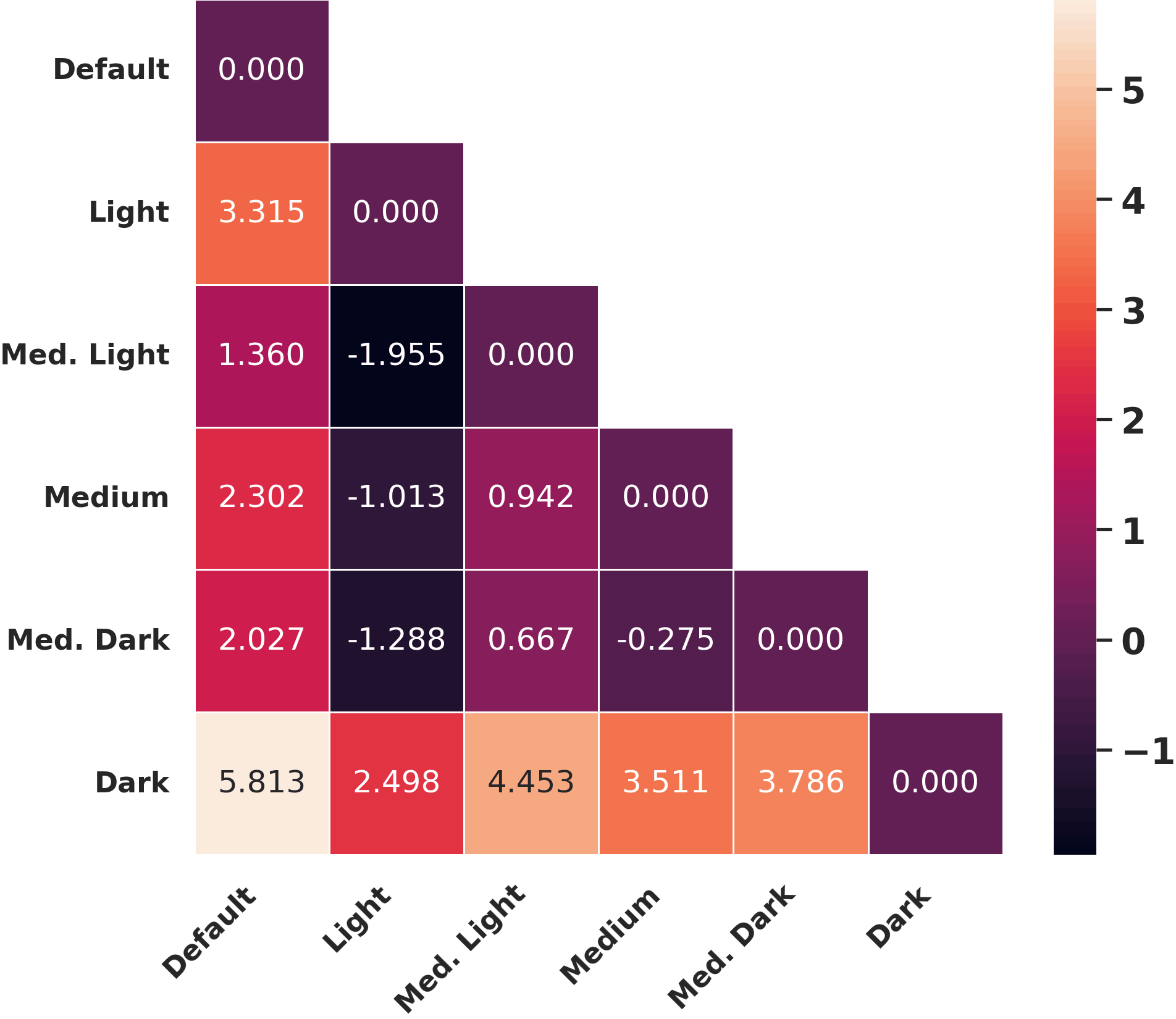}
        \label{fig:rnd_heatmap_gemma}
    }
    \subfigure[\llamaThreeTwoThreeB]{
        \includegraphics[width=0.18\linewidth]{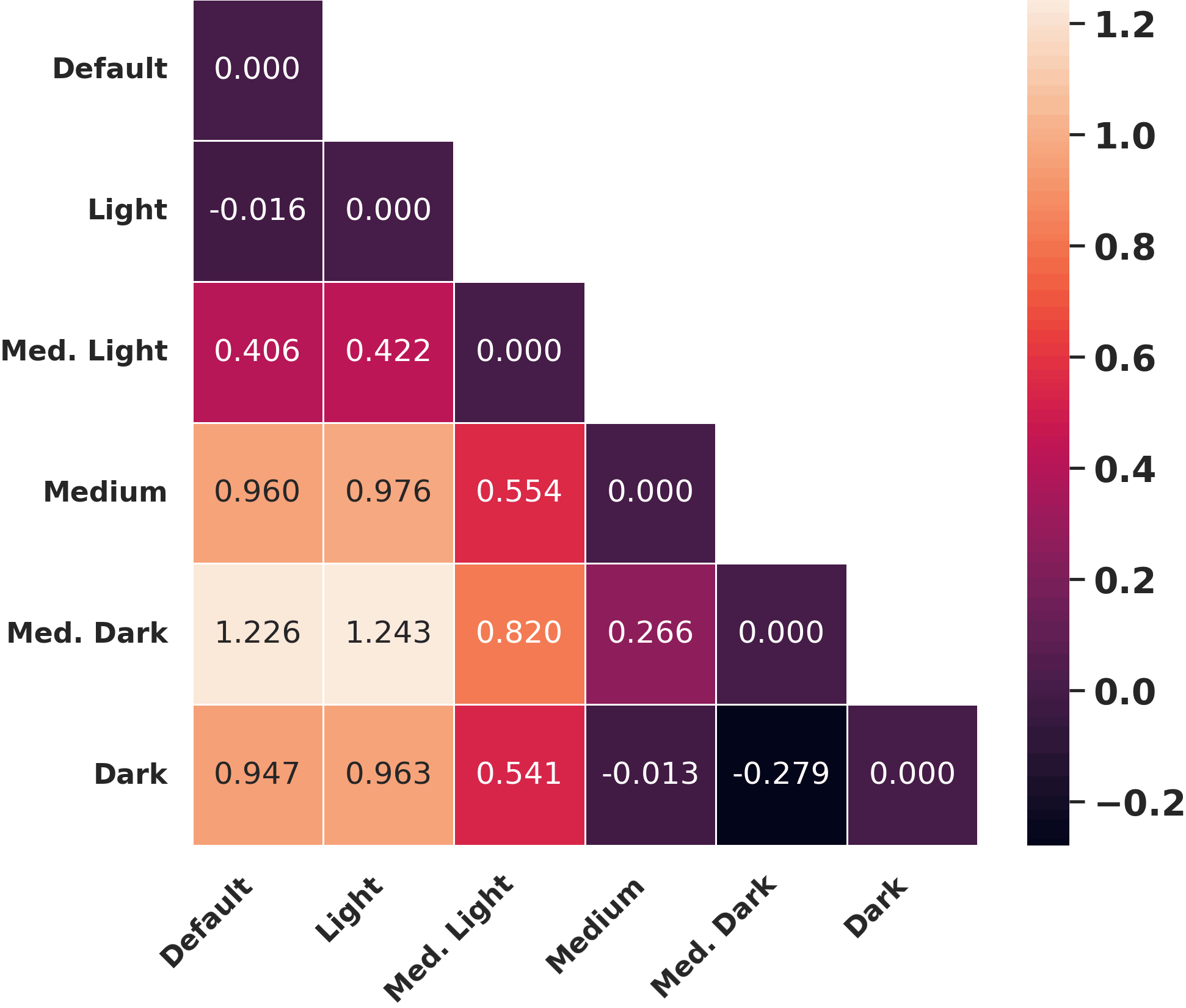}
        \label{fig:rnd_heatmap_llama}
    }
    \subfigure[\mistralVZeroThreeSevenB]{
        \includegraphics[width=0.18\linewidth]{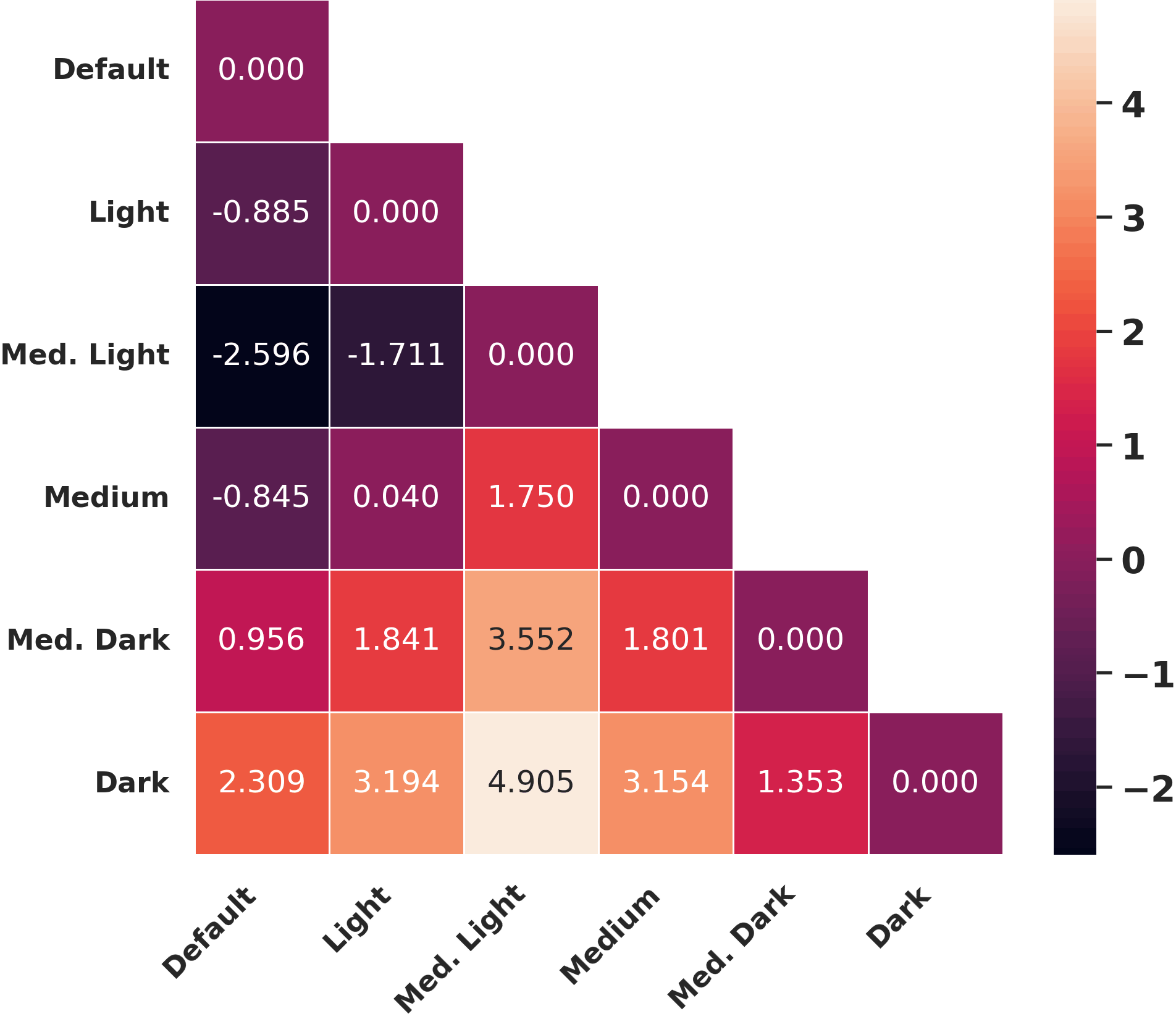}
        \label{fig:rnd_heatmap_mistral}
    }
    \subfigure[\qwenTwoSevenB]{
        \includegraphics[width=0.18\linewidth]{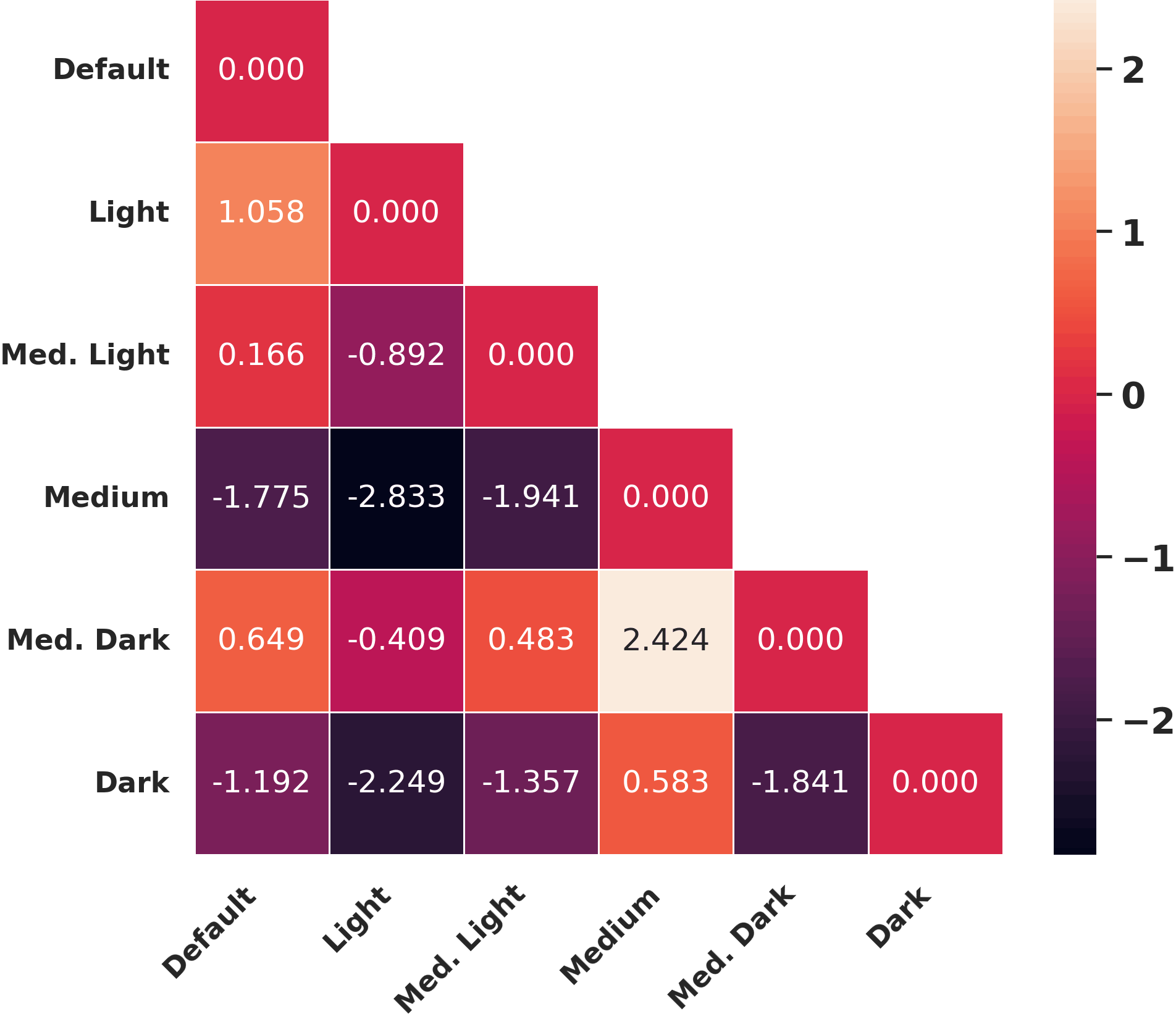}
        \label{fig:rnd_heatmap_qwen}
    }
    \vspace{-0.5em}
    \caption{Heatmaps illustrating the pairwise Relative Norm Distance (RND) scores for the hand gestures emoji subdomain across different embedding models. Each cell represents the RND score between two skin tone groups, where color intensity indicates the magnitude of the bias relative to a set of neutral words.}
    \label{fig:rnd_heatmaps_all}
\end{figure*}

The results in~\Cref{fig:rnd_heatmaps_all} reveal distinct patterns of neutral words bias across the models. The static emoji2vec model demonstrates robustness, maintaining a stable relationship with the neutral word set across different skin tones. In contrast, the LLMs exhibit larger and more varied biases. Among them, Gemma and Mistral are particularly noteworthy, as both models show a significant bias shift specifically involving the dark skin tone.

\subsection{Associative and Sentiment Bias in Person‑Role Emojis}
\label{sec:exp-person-role}
In this section, we examine whether skin tone variations on person role emojis create biased associations with sentiment. We utilize two complementary metrics: the Word Embedding Association Test (WEAT) and Relative Negative Sentiment Bias (RNSB). WEAT measures the implicit association between skin tone groups and predefined ``Good'' ([\includegraphics[width=0.02\linewidth]{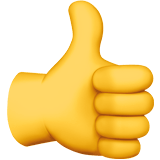}, \includegraphics[width=0.02\linewidth]{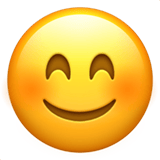}, \includegraphics[width=0.02\linewidth]{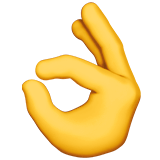}, \includegraphics[width=0.02\linewidth]{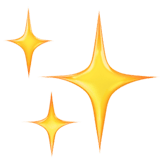}, \includegraphics[width=0.02\linewidth]{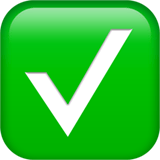}]) vs. ``Bad'' ([\includegraphics[width=0.02\linewidth]{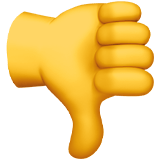}, \includegraphics[width=0.02\linewidth]{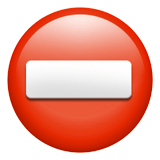}, \includegraphics[width=0.02\linewidth]{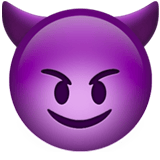}, \includegraphics[width=0.02\linewidth]{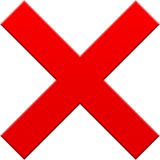}, \includegraphics[width=0.02\linewidth]{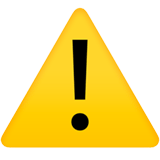}]) emoji \textbf{attribute} sets, which are sourced from the top five most relevant emojis on Emojidb\footnote{https://emojidb.org}. RNSB quantifies how unevenly negative sentiment is distributed among the variants of a single profession. This analysis uses emojis from the \textit{person role} category. A significant WEAT effect size or a high RNSB score would provide evidence of sentiment-based representational harms.

The WEAT analysis in~\Cref{tab:weat462_roles} reveals clear and model-dependent associative biases. When comparing \textit{light vs. dark} skin tones across professional roles, both Llama and Qwen exhibit consistently positive values. This indicates a shared tendency in these models to associate lighter skin tones more strongly with the ``Good'' emoji attribute set. In stark contrast, the Mistral model displays the opposite behavior, consistently yielding negative scores for the comparisons, including med. light and med. dark. This suggests an inherent bias within Mistral, where it has learned to associate lighter-skinned professional roles more closely with ``Bad'' attribute set.

While WEAT measures the direction of association, RNSB from~\Cref{tab:rnsb462_roles} allows us to quantify how unevenly negative sentiment is \textit{distributed} across the variants. The RNSB results confirm that this distribution is skewed, again with strong differences between models. Qwen exhibits the most uneven distribution of negative sentiment (highest KL divergence), followed by Llama, whereas Gemma's representations are considerably more balanced. Specifically, for Qwen, the \textit{default} and \textit{dark} skin tone variants consistently shoulder a disproportionately larger share of negative sentiment than mid-tones.

\begin{table}[h]
  \centering
  \caption{WEAT score across person roles. Cells report the mean effect size \(d\) per tone pair after averaging across all roles for each model. The row \textit{n (roles)} lists the number of available roles per model. For a tone pair \textit{X vs Y}, positive values mean \textit{X} is more associated with the ``Good'' emoji set (and \textit{Y} with ``Bad''); negative values mean \textit{X} is more associated with ``Bad'' (and \textit{Y} with ``Good'').}
  \label{tab:weat462_roles}
  \resizebox{\linewidth}{!}{
  \begin{tabular}{lrrrrr}
    \toprule
    \makecell[l]{\textbf{Tone Pair}\\(X vs. Y)} & \textbf{Emoji2vec} & \makecell[r]{\textbf{Gemma} \\ \scriptsize \gemmaTwoTwoB \normalsize} & \makecell[r]{\textbf{Llama} \\ \scriptsize \llamaThreeTwoThreeB \normalsize} & \makecell[r]{\textbf{Mistral} \\ \scriptsize \mistralVZeroThreeSevenB \normalsize} & \makecell[r]{\textbf{Qwen} \\ \scriptsize \qwenTwoSevenB \normalsize} \\
    \midrule
    n (roles) & 7 & 82 & 82 & 82 & 82 \\
    \midrule
    default vs \includegraphics[width=0.04\linewidth]{figures/emoji_image/light.jpg} & -0.286 & -0.537 & -1.415 & -0.634 & -0.829 \\
    default vs \includegraphics[width=0.04\linewidth]{figures/emoji_image/med_light.jpg} & -2.000 & -0.780 & -1.268 & -1.512 & -0.683 \\
    default vs \includegraphics[width=0.04\linewidth]{figures/emoji_image/med.jpg} & -1.429 & -1.073 & -1.122 & -1.707 & +1.220 \\
    default vs \includegraphics[width=0.04\linewidth]{figures/emoji_image/med_dark.jpg} & -0.286 & +0.244 & -1.024 & -1.561 & +0.439 \\
    default vs \includegraphics[width=0.04\linewidth]{figures/emoji_image/dark.jpg} & +0.286 & +1.073 & -0.537 & -1.902 & +1.220 \\
    \includegraphics[width=0.04\linewidth]{figures/emoji_image/light.jpg} vs \includegraphics[width=0.04\linewidth]{figures/emoji_image/med_light.jpg} & -2.000 & -0.341 & -0.439 & -1.220 & +0.146 \\
    \includegraphics[width=0.04\linewidth]{figures/emoji_image/light.jpg} vs \includegraphics[width=0.04\linewidth]{figures/emoji_image/med.jpg} & -1.429 & -0.439 & +0.927 & -1.756 & +1.317 \\
    \includegraphics[width=0.04\linewidth]{figures/emoji_image/light.jpg} vs \includegraphics[width=0.04\linewidth]{figures/emoji_image/med_dark.jpg} & +0.857 & +0.585 & +1.024 & -1.366 & +1.512 \\
    \includegraphics[width=0.04\linewidth]{figures/emoji_image/light.jpg} vs \includegraphics[width=0.04\linewidth]{figures/emoji_image/dark.jpg} & -0.286 & +1.610 & +1.220 & -1.463 & +1.659 \\
    \includegraphics[width=0.04\linewidth]{figures/emoji_image/med_light.jpg} vs \includegraphics[width=0.04\linewidth]{figures/emoji_image/med.jpg} & +0.857 & -0.439 & +0.829 & -1.268 & +1.366 \\
    \includegraphics[width=0.04\linewidth]{figures/emoji_image/med_light.jpg} vs \includegraphics[width=0.04\linewidth]{figures/emoji_image/med_dark.jpg} & +2.000 & +1.171 & +1.171 & -1.171 & +1.512 \\
    \includegraphics[width=0.04\linewidth]{figures/emoji_image/med_light.jpg} vs \includegraphics[width=0.04\linewidth]{figures/emoji_image/dark.jpg} & +2.000 & +1.707 & +1.122 & -1.610 & +1.463 \\
    \includegraphics[width=0.04\linewidth]{figures/emoji_image/med.jpg} vs \includegraphics[width=0.04\linewidth]{figures/emoji_image/med_dark.jpg} & +0.857 & +1.415 & +0.488 & -0.390 & -0.146 \\
    \includegraphics[width=0.04\linewidth]{figures/emoji_image/med.jpg} vs \includegraphics[width=0.04\linewidth]{figures/emoji_image/dark.jpg} & +1.429 & +1.902 & +0.878 & -1.122 & +1.415 \\
    \includegraphics[width=0.04\linewidth]{figures/emoji_image/med_dark.jpg} vs \includegraphics[width=0.04\linewidth]{figures/emoji_image/dark.jpg} & -0.857 & +0.878 & +0.927 & -1.171 & +1.024 \\
    \bottomrule
  \end{tabular}
  }
\end{table}

\begin{table}[h]
  \centering
  \caption{RNSB across person roles. Cells report across-role average negative share per tone; \textit{Avg KL} summarizes average KL divergence (higher indicates more uneven distributions).}
  \label{tab:rnsb462_roles}
  \resizebox{\linewidth}{!}{
  \begin{tabular}{lrrrrrrr}
    \toprule
    \textbf{Model} & \textbf{default} & \includegraphics[width=0.04\linewidth]{figures/emoji_image/light.jpg} & \includegraphics[width=0.04\linewidth]{figures/emoji_image/med_light.jpg} & \includegraphics[width=0.04\linewidth]{figures/emoji_image/med.jpg} & \includegraphics[width=0.04\linewidth]{figures/emoji_image/med_dark.jpg} & \includegraphics[width=0.04\linewidth]{figures/emoji_image/dark.jpg} & \textbf{Avg KL} \\
    \midrule
    emoji2vec & 0.418 & 0.167 & 0.160 & 0.165 & 0.169 & 0.170 & 0.0004 \\
    Gemma & 0.163 & 0.164 & 0.163 & 0.164 & 0.170 & 0.175 & 0.0057 \\
    Llama & 0.219 & 0.126 & 0.152 & 0.152 & 0.115 & 0.236 & 0.0821 \\
    Mistral & 0.127 & 0.155 & 0.177 & 0.157 & 0.180 & 0.204 & 0.0308 \\
    Qwen & 0.357 & 0.071 & 0.061 & 0.116 & 0.099 & 0.296 & 0.5310 \\
    \bottomrule
  \end{tabular}
  }
\end{table}

\subsection{Skin‑Tone Attributes vs Caliskan WEAT Benchmarks}
\label{sec:exp-WEAT}
We finally investigate whether emoji skin tone representations perpetuate broader societal stereotypes by applying the Word Embedding Association Test to a standard benchmark. In this analysis, we treat the skin tones themselves as the \textbf{target} sets, partitioning all skin-toned emojis into distinct groups (e.g., light-toned vs. dark-toned). We then measure the association between these emoji sets and the standard \textbf{attribute} word sets from the Caliskan et al. benchmark~\cite{caliskan2017semantics}, which cover established societal domains such as gender-career, race, and age. A statistically significant WEAT score would indicate that a model's representations have learned to link specific skin tones to harmful societal concepts.

\begin{table}[h]
  \centering
  \caption{WEAT results for the Caliskan et al. benchmark~\cite{caliskan2017semantics}, treating skin tones as targets. Cells report the mean effect size for each tone pair, averaged across all standard benchmarks for a given model. For a tone pair \textit{X vs Y}, a positive value indicates that skin tone \textit{X} is more strongly associated with culturally positive concepts. A negative value indicates that \textit{X} is more associated with the opponent concepts.}
  \label{tab:weat463_agg}
  \resizebox{\linewidth}{!}{
  \begin{tabular}{lrrrrr}
    \toprule
    \makecell[l]{\textbf{Tone Pair}\\(X vs. Y)} & \textbf{Emoji2vec} & \makecell[r]{\textbf{Gemma} \\ \scriptsize \gemmaTwoTwoB \normalsize} & \makecell[r]{\textbf{Llama} \\ \scriptsize \llamaThreeTwoThreeB \normalsize} & \makecell[r]{\textbf{Mistral} \\ \scriptsize \mistralVZeroThreeSevenB \normalsize} & \makecell[r]{\textbf{Qwen} \\ \scriptsize \qwenTwoSevenB \normalsize} \\
    \midrule
    default vs \includegraphics[width=0.04\linewidth]{figures/emoji_image/light.jpg} & -0.114 &  0.188 &  0.196 &  0.066 &  0.148 \\
    default vs \includegraphics[width=0.04\linewidth]{figures/emoji_image/med_light.jpg} & -0.004 &  0.121 & -0.080 &  0.253 &  0.143 \\
    default vs \includegraphics[width=0.04\linewidth]{figures/emoji_image/med.jpg} &  0.399 &  0.191 &  0.045 &  0.128 &  0.159 \\
    default vs \includegraphics[width=0.04\linewidth]{figures/emoji_image/med_dark.jpg} &  0.038 &  0.096 &  0.055 &  0.119 &  0.116 \\
    default vs \includegraphics[width=0.04\linewidth]{figures/emoji_image/dark.jpg} &  0.001 & -0.071 & -0.216 &  0.233 &  0.239 \\
    \includegraphics[width=0.04\linewidth]{figures/emoji_image/light.jpg} vs \includegraphics[width=0.04\linewidth]{figures/emoji_image/med_light.jpg} & -0.054 & -0.024 & -0.512 &  0.455 &  0.090 \\
    \includegraphics[width=0.04\linewidth]{figures/emoji_image/light.jpg} vs \includegraphics[width=0.04\linewidth]{figures/emoji_image/med.jpg} &  0.337 &  0.019 & -0.346 &  0.189 &  0.124 \\
    \includegraphics[width=0.04\linewidth]{figures/emoji_image/light.jpg} vs \includegraphics[width=0.04\linewidth]{figures/emoji_image/med_dark.jpg} &  0.043 & -0.035 & -0.385 &  0.114 &  0.036 \\
    \includegraphics[width=0.04\linewidth]{figures/emoji_image/light.jpg} vs \includegraphics[width=0.04\linewidth]{figures/emoji_image/dark.jpg} &  0.407 & -0.296 & -0.465 &  0.203 &  0.243 \\
    \includegraphics[width=0.04\linewidth]{figures/emoji_image/med_light.jpg} vs \includegraphics[width=0.04\linewidth]{figures/emoji_image/med.jpg} &  0.407 & -0.007 &  0.380 & -0.210 &  0.046 \\
    \includegraphics[width=0.04\linewidth]{figures/emoji_image/med_light.jpg} vs \includegraphics[width=0.04\linewidth]{figures/emoji_image/med_dark.jpg} &  0.112 & -0.010 &  0.415 & -0.119 & -0.049 \\
    \includegraphics[width=0.04\linewidth]{figures/emoji_image/med_light.jpg} vs \includegraphics[width=0.04\linewidth]{figures/emoji_image/dark.jpg} &  0.276 & -0.166 & -0.290 &  0.047 &  0.248 \\
    \includegraphics[width=0.04\linewidth]{figures/emoji_image/med.jpg} vs \includegraphics[width=0.04\linewidth]{figures/emoji_image/med_dark.jpg} & -0.303 & -0.056 &  0.050 &  0.006 & -0.178 \\
    \includegraphics[width=0.04\linewidth]{figures/emoji_image/med.jpg} vs \includegraphics[width=0.04\linewidth]{figures/emoji_image/dark.jpg} & -0.142 & -0.207 & -0.410 &  0.178 &  0.339 \\
    \includegraphics[width=0.04\linewidth]{figures/emoji_image/med_dark.jpg} vs \includegraphics[width=0.04\linewidth]{figures/emoji_image/dark.jpg} &  0.112 & -0.166 & -0.408 &  0.320 &  0.412 \\
    \bottomrule
  \end{tabular}
  }
  \vspace{-1em}
\end{table}

Our results, shown in~\Cref{tab:weat463_agg}, are interpreted based on the WEAT design principle, where attribute set A generally represents more culturally positive concepts than set B (see Appendix~\ref{sec:appendix_wefe_attr} for details). A clear pattern emerges where emoji2vec, Mistral, and Qwen exhibit a consistent preference for lighter skin tones. These models systematically associate relatively lighter tones with the more positive concepts in attribute set A, a tendency that holds even in comparisons between tones with minor differences. Conversely, Gemma and Llama display the opposite bias, consistently associating darker skin tones with the positive attribute sets.

\section{Conclusion}
\label{sec:Conclusion}
In this work, we conduct the first large-scale comparative study of skin-toned emoji representations, analyzing biases across both legacy static embedding models and modern LLMs. Our findings reveal a critical divide: while LLMs offer comprehensive support for the full spectrum of skin-toned emojis where older models fail, this support does not guarantee equitable representation. We empirically demonstrated that skin tone modifiers are not neutral but introduce measurable associative and sentiment-based biases into emoji embeddings, reinforcing harmful societal stereotypes. As LLMs become the foundational infrastructure for the web, these representational harms risk perpetuating inequities at a global scale. This underscores the urgent need to move beyond simple support to active mitigation, specifically through tokenizer normalization and counterfactual data augmentation, ensuring symbols of human identity are represented with genuine equity.

\bibliographystyle{ACM-Reference-Format}
\bibliography{WWW/www_bib}

@article{feng2020new,
  title={New emoji requests from Twitter users: when, where, why, and what we can do about them},
  author={Feng, Yunhe and Lu, Zheng and Zhou, Wenjun and Wang, Zhibo and Cao, Qing},
  journal={ACM Transactions on Social Computing},
  volume={3},
  number={2},
  pages={1--25},
  year={2020},
  publisher={ACM New York, NY, USA}
}

@inproceedings{hu2017spice,
  title={Spice up your chat: the intentions and sentiment effects of using emojis},
  author={Hu, Tianran and Guo, Han and Sun, Hao and Nguyen, Thuy-vy Thi and Luo, Jiebo},
  booktitle={Eleventh international aaai conference on web and social media},
  year={2017}
}

@inproceedings{barbieri2018gender,
  title={How gender and skin tone modifiers affect emoji semantics in twitter},
  author={Barbieri, Francesco and Camacho-Collados, Jose},
  year={2018},
  organization={The Association for Computational Linguistics}
}

@article{reelfs2020word,
  title={Word-emoji embeddings from large scale messaging data reflect real-world semantic associations of expressive icons},
  author={Reelfs, Jens Helge and Hohlfeld, Oliver and Strohmaier, Markus and Henckell, Niklas},
  journal={arXiv preprint arXiv:2006.01207},
  year={2020}
}

@inproceedings{barry2021emojional,
  title={Emojional: Emoji Embeddings},
  author={Barry, Elena and Jameel, Shoaib and Raza, Haider},
  booktitle={UK Workshop on Computational Intelligence},
  pages={312--324},
  year={2021},
  organization={Springer}
}

@article{mikolov2013efficient,
  title={Efficient estimation of word representations in vector space},
  author={Mikolov, Tomas and Chen, Kai and Corrado, Greg and Dean, Jeffrey},
  journal={arXiv preprint arXiv:1301.3781},
  year={2013}
}

@misc{AdiShirsath2021,
  author = {Aditya Shirsath},
  title = {Training Word2Vec model for emojis using twitter data},
  year = {2021},
  publisher = {GitHub},
  journal = {GitHub repository},
  howpublished = {\url{https://github.com/AdiShirsath/Emoji_Word2Vec}}
}

@article{garg2018word,
  title={Word embeddings quantify 100 years of gender and ethnic stereotypes},
  author={Garg, Nikhil and Schiebinger, Londa and Jurafsky, Dan and Zou, James},
  journal={Proceedings of the National Academy of Sciences},
  volume={115},
  number={16},
  pages={E3635--E3644},
  year={2018},
  publisher={National Acad Sciences}
}

@article{fitzpatrick1988validity,
  title={The validity and practicality of sun-reactive skin types I through VI},
  author={Fitzpatrick, Thomas B},
  journal={Archives of dermatology},
  volume={124},
  number={6},
  pages={869--871},
  year={1988},
  publisher={American Medical Association}
}

@article{eisner2016emoji2vec,
  title={emoji2vec: Learning emoji representations from their description},
  author={Eisner, Ben and Rockt{\"a}schel, Tim and Augenstein, Isabelle and Bo{\v{s}}njak, Matko and Riedel, Sebastian},
  journal={arXiv preprint arXiv:1609.08359},
  year={2016}
}

@article{bojanowski2016enriching,
  title={Enriching Word Vectors with Subword Information},
  author={Bojanowski, Piotr and Grave, Edouard and Joulin, Armand and Mikolov, Tomas},
  journal={arXiv preprint arXiv:1607.04606},
  year={2016}
}

@inproceedings{pennington2014glove,
  title={Glove: Global vectors for word representation},
  author={Pennington, Jeffrey and Socher, Richard and Manning, Christopher D},
  booktitle={Proceedings of the 2014 conference on empirical methods in natural language processing (EMNLP)},
  pages={1532--1543},
  year={2014}
}

@inproceedings{sweeney2019transparent,
  title={A transparent framework for evaluating unintended demographic bias in word embeddings},
  author={Sweeney, Chris and Najafian, Maryam},
  booktitle={Proceedings of the 57th Annual Meeting of the Association for Computational Linguistics},
  pages={1662--1667},
  year={2019}
}

@article{devlin2018bert,
  title={Bert: Pre-training of deep bidirectional transformers for language understanding},
  author={Devlin, Jacob and Chang, Ming-Wei and Lee, Kenton and Toutanova, Kristina},
  journal={arXiv preprint arXiv:1810.04805},
  year={2018}
}

@article{kaye2016turn,
  title={“Turn that frown upside-down”: A contextual account of emoticon usage on different virtual platforms},
  author={Kaye, Linda K and Wall, Helen J and Malone, Stephanie A},
  journal={Computers in Human Behavior},
  volume={60},
  pages={463--467},
  year={2016},
  publisher={Elsevier}
}

@article{hovy2021five,
  title={Five sources of bias in natural language processing},
  author={Hovy, Dirk and Prabhumoye, Shrimai},
  journal={Language and Linguistics Compass},
  volume={15},
  number={8},
  pages={e12432},
  year={2021},
  publisher={Wiley Online Library}
}

@article{caliskan2017semantics,
  title={Semantics derived automatically from language corpora contain human-like biases},
  author={Caliskan, Aylin and Bryson, Joanna J and Narayanan, Arvind},
  journal={Science},
  volume={356},
  number={6334},
  pages={183--186},
  year={2017},
  publisher={American Association for the Advancement of Science}
}

@inproceedings{wiseman2018repurposing,
  title={Repurposing emoji for personalised communication: Why the pizza emoji means “I love you”},
  author={Wiseman, Sarah and Gould, Sandy JJ},
  booktitle={Proceedings of the 2018 CHI conference on human factors in computing systems},
  pages={1--10},
  year={2018}
}

@article{santhanam2019stand,
  title={I stand with you: Using emojis to study solidarity in crisis events},
  author={Santhanam, Sashank and Srinivasan, Vidhushini and Glass, Shaina and Shaikh, Samira},
  journal={arXiv preprint arXiv:1907.08326},
  year={2019}
}

@article{kralj2015sentiment,
  title={Sentiment of emojis},
  author={Kralj Novak, Petra and Smailovi{\'c}, Jasmina and Sluban, Borut and Mozeti{\v{c}}, Igor},
  journal={PloS one},
  volume={10},
  number={12},
  pages={e0144296},
  year={2015},
  publisher={Public Library of Science San Francisco, CA USA}
}

@inproceedings{tigwell2016oh,
  title={Oh that's what you meant! Reducing emoji misunderstanding},
  author={Tigwell, Garreth W and Flatla, David R},
  booktitle={Proceedings of the 18th international conference on human-computer interaction with mobile devices and services adjunct},
  pages={859--866},
  year={2016}
}

@inproceedings{miller2017understanding,
  title={Understanding emoji ambiguity in context: The role of text in emoji-related miscommunication},
  author={Miller, Hannah and Kluver, Daniel and Thebault-Spieker, Jacob and Terveen, Loren and Hecht, Brent},
  booktitle={Proceedings of the International AAAI Conference on Web and Social Media},
  volume={11},
  number={1},
  pages={152--161},
  year={2017}
}

@book{perez2019invisible,
  title={Invisible women: Data bias in a world designed for men},
  author={Perez, Caroline Criado},
  year={2019},
  publisher={Abrams}
}

@inproceedings{barbosa2019rehumanized,
  title={Rehumanized crowdsourcing: A labeling framework addressing bias and ethics in machine learning},
  author={Barbosa, Nat{\~a} M and Chen, Monchu},
  booktitle={Proceedings of the 2019 CHI Conference on Human Factors in Computing Systems},
  pages={1--12},
  year={2019}
}

@incollection{noble2018algorithms,
  title={Algorithms of oppression},
  author={Noble, Safiya Umoja},
  booktitle={Algorithms of oppression},
  year={2018},
  publisher={New York university press}
}

@inproceedings{raji2019actionable,
  title={Actionable auditing: Investigating the impact of publicly naming biased performance results of commercial ai products},
  author={Raji, Inioluwa Deborah and Buolamwini, Joy},
  booktitle={Proceedings of the 2019 AAAI/ACM Conference on AI, Ethics, and Society},
  pages={429--435},
  year={2019}
}

@article{brock2011keeping,
  title={‘‘When Keeping it Real Goes Wrong’’: Resident Evil 5, Racial Representation, and Gamers},
  author={Brock, Andr{\'e}},
  journal={Games and Culture},
  volume={6},
  number={5},
  pages={429--452},
  year={2011},
  publisher={Sage Publications Sage CA: Los Angeles, CA}
}

@article{cave2020whiteness,
  title={The whiteness of AI},
  author={Cave, Stephen and Dihal, Kanta},
  journal={Philosophy \& Technology},
  volume={33},
  number={4},
  pages={685--703},
  year={2020},
  publisher={Springer}
}

@article{sparrow2020robotics,
  title={Robotics has a race problem},
  author={Sparrow, Robert},
  journal={Science, Technology, \& Human Values},
  volume={45},
  number={3},
  pages={538--560},
  year={2020},
  publisher={SAGE Publications Sage CA: Los Angeles, CA}
}

@article{brock2011beyond,
  title={Beyond the pale: The Blackbird web browser’s critical reception},
  author={Brock, Andr{\'e}},
  journal={New Media \& Society},
  volume={13},
  number={7},
  pages={1085--1103},
  year={2011},
  publisher={Sage Publications Sage UK: London, England}
}

@article{zimmerman2015racially,
  title={Racially diverse emoji are a nice idea. But will anyone use them},
  author={Zimmerman, J},
  journal={The Guardian},
  year={2015}
}

@article{miltner2021one,
  title={“One part politics, one part technology, one part history”: Racial representation in the Unicode 7.0 emoji set},
  author={Miltner, Kate M},
  journal={New Media \& Society},
  volume={23},
  number={3},
  pages={515--534},
  year={2021},
  publisher={SAGE Publications Sage UK: London, England}
}

@MISC{Arielle2023,
author = {Arielle Pardes},
title = {The solution to the emoji diversity problem: Make them all yellow},
month = {Septmber},
year = {2015},
howpublished={\url{https://www.vice.com/en/article/wd7ejm/emoji-shouldve-made-all-their-characters-yellow-408}}
}

@misc{mistral7b_instruct_v0.3,
  title        = {Mistral-7B-Instruct-v0.3},
  author       = {Mistral AI},
  year         = {2024},
  publisher    = {Hugging Face},
  howpublished = {\url{https://huggingface.co/mistralai/Mistral-7B-Instruct-v0.3}},
  note         = {Apache 2.0 License}
}

@article{qwen2,
  title={Qwen2 Technical Report},
  author={Team, Qwen and others},
  year={2024}
}

@misc{meta_llama3.2_3b_instruct,
  title        = {Meta-Llama-3.2-3B-Instruct},
  author       = {Meta},
  year         = {2024},
  publisher    = {Hugging Face},
  howpublished = {\url{https://huggingface.co/meta-llama/Llama-3.2-3B-Instruct}},
  note         = {Llama 3.2 Community License}
}

@misc{meta_llama3.2_1b_instruct,
  title        = {Meta-Llama-3.2-1B-Instruct},
  author       = {Meta},
  year         = {2024},
  publisher    = {Hugging Face},
  howpublished = {\url{https://huggingface.co/meta-llama/Llama-3.2-1B-Instruct}},
  note         = {Llama 3.2 Community License}
}

@article{gemma_2024,
    title={Gemma},
    url={https://www.kaggle.com/m/3301},
    DOI={10.34740/KAGGLE/M/3301},
    publisher={Kaggle},
    author={Gemma Team},
    year={2024}
}

@article{sennrich2015neural,
  title={Neural machine translation of rare words with subword units},
  author={Sennrich, Rico and Haddow, Barry and Birch, Alexandra},
  journal={arXiv preprint arXiv:1508.07909},
  year={2015}
}

@article{kudo2018sentencepiece,
  title={SentencePiece: A simple and language independent subword tokenizer and detokenizer for neural text processing},
  author={Kudo, Taku and Richardson, John},
  journal={arXiv preprint arXiv:1808.06226},
  year={2018}
}

@article{radford2018improving,
  title={Improving language understanding by generative pre-training},
  author={Radford, Alec and Narasimhan, Karthik and Salimans, Tim and Sutskever, Ilya and others},
  year={2018},
  publisher={San Francisco, CA, USA}
}

@inproceedings{mohammad2018obtaining,
  title={Obtaining reliable human ratings of valence, arousal, and dominance for 20,000 English words},
  author={Mohammad, Saif},
  booktitle={Proceedings of the 56th annual meeting of the association for computational linguistics (volume 1: Long papers)},
  pages={174--184},
  year={2018}
}

@inproceedings{kusner2015word,
  title={From word embeddings to document distances},
  author={Kusner, Matt and Sun, Yu and Kolkin, Nicholas and Weinberger, Kilian},
  booktitle={International conference on machine learning},
  pages={957--966},
  year={2015},
  organization={PMLR}
}

@article{vaswani2017attention,
  title={Attention is all you need},
  author={Vaswani, Ashish and Shazeer, Noam and Parmar, Niki and Uszkoreit, Jakob and Jones, Llion and Gomez, Aidan N and Kaiser, {\L}ukasz and Polosukhin, Illia},
  journal={Advances in neural information processing systems},
  volume={30},
  year={2017}
}

\appendix

\section{Detailed Token Distribution Analysis}
\label{sec:appendix_token_dist}

\begin{figure}[h]
    \centering
    \includegraphics[width=0.7\linewidth]{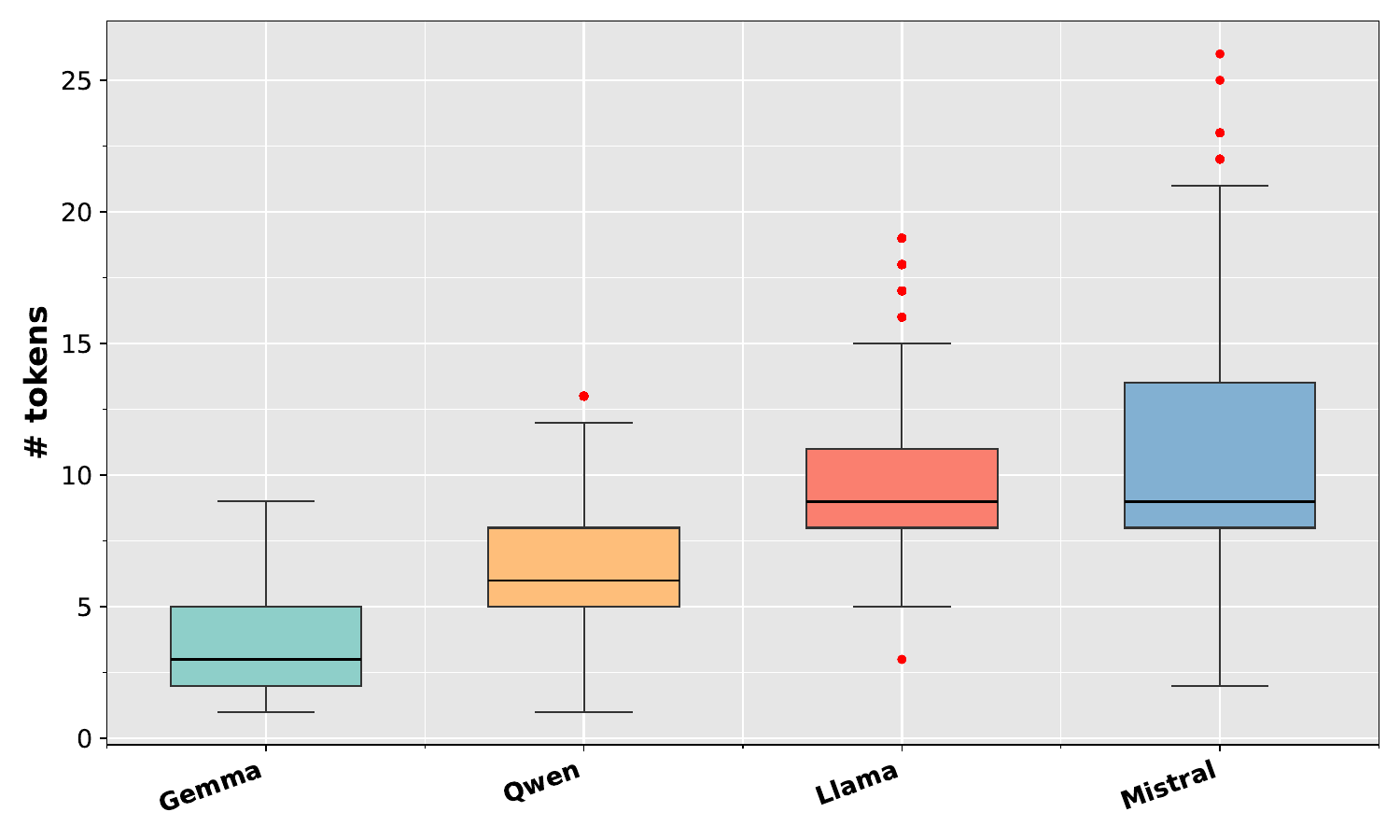}
    \caption{A boxplot comparing the distribution of the number of tokens per skin-toned emoji across different LLMs. Each box shows the median, quartiles, and range, illustrating the variance in tokenization strategies.}
    \label{fig:token_boxplot}
\end{figure}

To complement the summary statistics in the main text, Figure~\ref{fig:token_boxplot} presents a boxplot visualizing the distribution of token counts required to represent the full set of 2,735 skin-toned emojis across the five tested LLMs. This visualization highlights the differences in median token count, interquartile range (IQR), and the presence of outliers for each model, offering a clearer picture of their tokenization efficiency and consistency.

\section{Attribute set for WEFE}
\label{sec:appendix_wefe_attr}
We construct our analysis by adopting the standard attribute and target pairs from the Caliskan et al. WEAT benchmark~\cite{caliskan2017semantics}. This design follows a consistent principle where Attribute A represents concepts with more culturally positive or stereotypically favored connotations, while Attribute B represents their conceptual opposites. For instance, stereotypically pleasant items like \textit{flowers} are contrasted with \textit{insects}, and creative tools like \textit{instruments} are contrasted with harmful ones like \textit{weapons}. ~\Cref{tab:wefe_attr_samples} lists the full inventory of attribute pairs used in our study, with sample tokens to aid interpretation.

\begin{table}[h]
  \centering
  \caption{WEFE attribute pairs with two samples per attribute.}
  \label{tab:wefe_attr_samples}
  \scriptsize
  \setlength{\tabcolsep}{4pt}
  \begin{tabular}{llll}
    \toprule
    \textbf{Attribute A} & \textbf{Sample A} & \textbf{Attribute B} & \textbf{Sample B} \\
    \midrule
    flowers & aster, clover & insects & ant, caterpillar \\
    instruments & bagpipe, cello & weapons & arrow, club \\
    \makecell[l]{european american\\ names 7} & Brad, Brendan & \makecell[l]{african american\\ names 7} & Darnell, Hakim \\
    \makecell[l]{european american\\ names 5} & Adam, Harry & \makecell[l]{african american\\ names 5} & Alonzo, Jamel \\
    male names & John, Paul & female names & Amy, Joan \\
    math & math, algebra & arts & poetry, art \\
    science & science, technology & arts 2 & poetry, art \\
    mental disease & sad, hopeless & physical disease & sick, illness \\
    young people names & Tiffany, Michelle & old people names & Ethel, Bernice \\
    pleasant 5 & caress, freedom & unpleasant 5 & abuse, crash \\
    pleasant 9 & joy, love & unpleasant 9 & agony, terrible \\
    career & executive, management & family & home, parents \\
    male terms & male, man & female terms & female, woman \\
    male terms 2 & brother, father & female terms 2 & sister, mother \\
    temporary & impermanent, unstable & permanent & stable, always \\
    \bottomrule
  \end{tabular}
\end{table}

\section{Bias Statistical Significance Analysis}
\label{sec:appendix_sig}
We report, for each tone pair and model, the percentage of de-duplicated benchmarks reaching \(p<0.05\) under standard WEAT permutation testing (10k permutations, fixed seed). Higher percentages indicate that an effect is significant across more semantic axes.

\begin{table}[h]
  \centering
  \caption{Significance rate (\% of benchmarks with \(p<0.05\)) per tone pair and model.}
  \label{tab:weat463_sig_pct}
  \resizebox{\linewidth}{!}{
  \begin{tabular}{lrrrrr}
    \toprule
    \textbf{Tone Pair} & \textbf{emoji2vec} & \textbf{Gemma} & \textbf{Llama} & \textbf{Mistral} & \textbf{Qwen} \\
    \midrule
    default vs \includegraphics[width=0.04\linewidth]{figures/emoji_image/light.jpg} & 47\% &  7\% &  7\% & 13\% & 33\% \\
    default vs \includegraphics[width=0.04\linewidth]{figures/emoji_image/med_light.jpg} & 27\% & 27\% & 13\% & 20\% & 33\% \\
    default vs \includegraphics[width=0.04\linewidth]{figures/emoji_image/med.jpg} & 47\% & 13\% &  0\% & 27\% & 27\% \\
    default vs \includegraphics[width=0.04\linewidth]{figures/emoji_image/med_dark.jpg} & 40\% & 20\% &  0\% & 13\% & 27\% \\
    default vs \includegraphics[width=0.04\linewidth]{figures/emoji_image/dark.jpg} & 53\% & 20\% & 20\% &  7\% & 27\% \\
    \includegraphics[width=0.04\linewidth]{figures/emoji_image/light.jpg} vs \includegraphics[width=0.04\linewidth]{figures/emoji_image/med_light.jpg} & 27\% & 20\% & 40\% & 60\% & 13\% \\
    \includegraphics[width=0.04\linewidth]{figures/emoji_image/light.jpg} vs \includegraphics[width=0.04\linewidth]{figures/emoji_image/med.jpg} & 40\% &  0\% & 20\% & 13\% & 33\% \\
    \includegraphics[width=0.04\linewidth]{figures/emoji_image/light.jpg} vs \includegraphics[width=0.04\linewidth]{figures/emoji_image/med_dark.jpg} & 13\% &  7\% & 33\% & 13\% & 20\% \\
    \includegraphics[width=0.04\linewidth]{figures/emoji_image/light.jpg} vs \includegraphics[width=0.04\linewidth]{figures/emoji_image/dark.jpg} & 40\% & 13\% & 33\% & 13\% & 33\% \\
    \includegraphics[width=0.04\linewidth]{figures/emoji_image/med_light.jpg} vs \includegraphics[width=0.04\linewidth]{figures/emoji_image/med.jpg} & 40\% &  0\% & 20\% &  7\% & 13\% \\
    \includegraphics[width=0.04\linewidth]{figures/emoji_image/med_light.jpg} vs \includegraphics[width=0.04\linewidth]{figures/emoji_image/med_dark.jpg} & 33\% &  0\% & 20\% &  0\% & 20\% \\
    \includegraphics[width=0.04\linewidth]{figures/emoji_image/med_light.jpg} vs \includegraphics[width=0.04\linewidth]{figures/emoji_image/dark.jpg} & 40\% & 13\% & 20\% & 13\% & 27\% \\
    \includegraphics[width=0.04\linewidth]{figures/emoji_image/med.jpg} vs \includegraphics[width=0.04\linewidth]{figures/emoji_image/med_dark.jpg} & 47\% &  7\% &  7\% & 13\% & 20\% \\
    \includegraphics[width=0.04\linewidth]{figures/emoji_image/med.jpg} vs \includegraphics[width=0.04\linewidth]{figures/emoji_image/dark.jpg} & 73\% & 13\% & 27\% & 27\% & 40\% \\
    \includegraphics[width=0.04\linewidth]{figures/emoji_image/med_dark.jpg} vs \includegraphics[width=0.04\linewidth]{figures/emoji_image/dark.jpg} & 27\% &  0\% & 27\% & 20\% & 27\% \\
    \bottomrule
  \end{tabular}
  }
\end{table}

On average, emoji2vec and Qwen exhibit the highest overall statistical significance coverage, while Gemma shows the lowest rates. However, this effect is not uniform, with significance rates varying widely depending on the specific tone pair being compared. A clear trend emerges where contrasts involving the dark skin tone yield more statistically significant results, indicating more stable learned associations. For example, the medium vs. dark comparison produces the highest significance rate in the entire dataset for emoji2vec (73\%) and a high rate for Qwen (40\%).

\end{document}